\documentclass[useAMS,usenatbib,usegraphicx]{mn2e}
\usepackage{graphicx}
\usepackage[colorlinks=true,citecolor=blue,linkcolor=blue]{hyperref}
\usepackage{amssymb,natbib}

\usepackage{fancyhdr}
\usepackage{longtable,lscape,amsmath}  
\usepackage{rotating}
\usepackage{xcolor}
\usepackage{natbib,twoopt}
\usepackage[T1]{fontenc}
\usepackage{aecompl}

\def\kmss    {km\, s$^{-1}$}
\def\kms     {km\, s$^{-1}$\,}

\include{definitions}

\begin{document}

\title[Formaldehyde Megamasers]{The Emission Structure of Formaldehyde MegaMasers}
\author[Baan et al.]
{
Willem A.~Baan$^{1,2}$,  Tao~An$^{2,1,3}$, Hans-Rainer Kl\"ockner$^{4}$, and Peter Thomasson$^{5}$
\\
$^1$Netherlands Institute for Radio Astronomy ASTRON, 7991 PD Dwingeloo, The Netherlands \\
$^2$Shanghai Astronomical Observatory, Chinese Academy of Sciences, 200030 Shanghai, China\\
$^3$Key Laboratory of Radio Astronomy, Chinese Academy of Sciences, 210008 Nanjing, China\\
$^4$Max Planck Institut f\"ur Radioastronomie, 69 Auf den H\"ugel, 5300 Bonn, Germany\\
$^5$University of Manchester, Jodrell Bank Observatory, Macclesfield, Cheshire SK11 9DL, UK\\
}

\date{Accepted xxx. Received xxx; in original form xxx}
\pagerange{\pageref{firstpage}--\pageref{lastpage}} \pubyear{2017}

\maketitle

\begin{abstract}
The formaldehyde MegaMaser emission has been mapped for the three host galaxies IC\,860. 
IRAS\,15107$+$0724, and Arp\,220. Elongated emission components are found at the nuclear 
centres of all galaxies with an extent ranging between 30 to 100 pc. These components are
superposed on the peaks of the nuclear continuum. Additional isolated emission components are found 
superposed in the outskirts of the radio continuum structure. The brightness temperatures
of the detected features ranges from 0.6 to 13.4 $\times 10^{4}$ K, which confirms their masering 
nature. The masering scenario is interpreted as amplification of the radio continuum by 
foreground molecular gas that is pumped by far-infrared radiation fields in these starburst environments of 
the host galaxies.
\end{abstract}


%

\section{Introduction}
\label{sec:introduction}

High-luminosity MegaMaser (MM) emission has been detected in external galaxies for four molecular species. 
Hydroxyl (OH) masers are found in the nuclear regions of Ultra-luminous Infrared Galaxies ((U)LIRGs) 
\citep{Baan82}, water vapour (H$_2$O) masers in accretion disks of active galactic nuclei (AGN) 
\citep{BraatzEA10, KuoEA15} and in interaction 
regions of jets and circumnuclear clouds \citep[e.g.,][]{Claussen84,HB85,Middelberg07}, 
formaldehyde (H$_2$CO) masers in the nuclear regions of ULIRGs \citep{Baan86, Araya04}, and recently methanol 
(CH$_3$OH) and Silicon Oxide (SiO) maser emission has been found in the starburst regions of (U)LIRGs as well as in nuclear feedback regions \citep{Ellingsen14,Chen15,WangEA14}.

A significant contribution to these observed molecular emissions may result from the amplification of the embedded or background radio continuum emission by foreground pumped molecular gas, as already proposed for the prototype OH MM source 
Arp\,220 (IC\,4553 - IRAS\,15234$+$2354) \citep{Baan82}.
Rather than relying on the conventional high-gain maser scenario with a tunnel-like column of inverted molecules that 
amplifies some spontaneous seed emission, a large fraction of these OH MM emissions is produced by low-gain unsaturated amplification of the background radio continuum \citep{Baan85}. This scenario naturally accounts for the superposition of maser emission 
originating in molecular cloud structures with variable local pumping conditions and the extended radio continuum in 
the source \citep{Baan89,Parra05}, resulting in both low- and high-brightness maser components. 


Superposition of maser and continuum sources has already clearly been shown for the powerful OH MM sources 
such as Arp\,220 \citep[e.g.,][]{Baan87}, Mrk\,273 \citep[e.g.,][]{Klockner04}, and IRAS\,17207$-$0054 \citep[e.g.,][]{Momjian06}, 
as well as for the strong central components of the H$_2$O MM source NGC\,4258 \citep{Miyoshi95,Herrnstein97} 
and other powerful sources such as NGC\,3079 {having an extended nuclear continuum structure} \citep[e.g.,][]{Haschick90, Kondratko05}. 
Both the Class I and II methanol MM emissions observed in nearby starburst galaxies also suggest that they are confined within the extended continuum emission \citep{Ellingsen14,Chen15}   
A first map of the formaldehyde emission in Arp\,220 also shows a superposition of line emission and continuum 
\citep{Baan95} and the current study should confirm this composite maser amplification scenario. 
Similarly, Galactic OH and H$_2$O maser sources lose much of their flux at long terrestrial and space interferometer baselines, indicating the contribution of extended low-brightness maser emission \citep{Slysh01}.

Formaldehyde masering activity remains very rare in the Galaxy and in extragalactic sources, and the 
observed masers are relatively weak and difficult to find. 	
Although formaldehyde absorption is widespread in the Galaxy, formaldehyde emission in the 4.829~GHz  
K$_a$ = 1$_{10}$$-$1$_{11}$ transition is currently known to occur in only eight Galactic sources 
\citep[e.g.,][]{Araya08,GinsburgEA15}. 
The H$_2$CO MM emission in the extragalactic sources IC\,860 and IRAS\,15107$+$0724, and Arp\,220 have 
been confirmed  \citep{Baan93,Araya04,Mangum08}, while other sources require interferometric confirmation. 
In addition, the source NGC\,6240 displays two emission components that probably coincide with lower brightness 
continuum structures and that cover the velocity range of the single-dish spectrum \citep{Baan93,WangEA14}.
While most extragalactic sources show absorption in both the 4.83~GHz K$_a$ = 1$_{10}$$-$1$_{11}$ and the 
14.5~GHz K$_a$ = 2$_{11}$$-$2$_{12}$ transitions \citep{Mangum13}, the few sources with ground-state emission 
also exhibit dominant absorption in the 14.5~GHz transition except for possible partial infilling of the line by 
emission. 
The current study suggests that the localised emission in the ground state may also be accompanied by more 
distributed absorption. The detection of such weak localised emission components embedded within more 
extended absorption requires higher resolution observations.

The masering activity in extragalactic sources strongly depends on the availability of an appropriate 
pumping agent for the molecular species and favourable environmental conditions. 
The known OH MM emissions generally occur in (U)LIRGs where radiative pumping of the OH results from 
star-formation induced dust infrared emissions \citep{Baan89,Henkel90,Darling02}. 
Since all three H$_2$CO MM sources are also known as OH MMs, it may be assumed that the 
infrared radiation fields in the nuclei of these galaxies also contribute to the pumping of the formaldehyde.

Issues to be raised in this study of IC\,860, IRAS\,15107$+$0724, and Arp\,220 are the spatial structure of the 
masering formaldehyde emission components, their superposition on the nuclear continuum emission and 
their velocity characteristics. The observed brightness temperature of these components and the radio continuum 
will be used confirm the masering nature of the line emission. 
This paper will also address the ability of the infrared radiation field to pump the masering activity of formaldehyde 
and to provide sufficient gains.  
A final issue to be considered is to seek evidence of more extended absorption in these sources. 

\section{MERLIN Observations}
\label{sec:observations}


The 4.83 GHz H$_2$CO K$_a$ = 1$_{10}$$-$1$_{11}$ transition data for the sources IC\,860 (IRAS\,13126$+$2452), 
IRAS\,15107$+$0724 and Arp\,220 (IRAS\,15234$+$2354) have been obtained with the MERLIN array using the
antennas at Defford, Cambridge, Knockin, Darnhall, Jodrell Mk2, Lovell, and Tabley.  

IC\,860 was observed from November 2009 for a total of approximately 35 hours including the flux density calibrator, 
3C\,286 (7.57 Jy), and the phase reference source, 1318+225 (0.26 Jy). 
The total target on-source time was 23.4 hrs, part of which was affected by weather. 
IRAS\,15107+0724 was also observed during November 2009 for a total of 36.7 hours including the flux 
calibrator, 3C\,286, baseline calibrator,  OQ\,208 (0.99 Jy), and phase reference source, 1509+054 (0.84 Jy). 
The target on-source time was 27.1 hrs, part of which was also affected by weather.
The observations of both IRAS\,15107+0724 and IC\,860 have an observing bandwidth of 16 MHz centred at a 
sky frequency of 4768.4 MHz, which corresponds to a redshift at band centre for both sources of 3802.8 \kmss. 
 
The heliocentric radial velocity of 3347 \kms for IC\,860 indicates a distance of 46.0 Mpc and a spatial scale of 
223 pc/arcsec. 
The emission from IC\,860 has been found to be close to its systemic velocity of 3911 \kms and not at the 
heliocentric velocity of 3347 \kms found in the literature. 
The heliocentric radial velocity of 3897 \kms for IRAS\,15107+0724 indicates a distance 53.8 Mpc and a spatial 
scale of 261 pc/arcsec.

\begin{table}
\begin{center}
\caption{The Continuum Components}
\label{table1}
\begin{tabular}{lcccc}
\hline
Source          &RA$^a$         & Dec$^a$ &Continuum            & Brightness       \\
Component           &               &                      & Flux             &   Temerature              \\
                       &  (s) & (\arcsec)   & (mJy/b)   &   (10$^4$ K) \\
\hline
                       &                &                   &                   &                                      \\
\multicolumn{2}{l}{IC\,860} & & &   \\
C           &  03.505 & 07.791 &4.68 & 5.13 \\
E           &  03.516 & 07.755  &0.61 & 0.45 \\
S           &  03.500 & 07.575  &0.78 & 0.63 \\
W          &  03.498  & 07.820 & 1.75   & 0.78 \\
NW        & 03.497  & 07.946  & 0.75 & 0.12  \\
                    &           &                       &                      &                       \\
\multicolumn{2}{l}{IRAS\,15107+0724} & & &   \\
C           & 13.094 &  31.872 & 7.92 &  9.40  \\
NW     & 13.091 &  32.030 & 0.72&  0.53 \\
W        & 13.083 &  31.847 &  0.47&  0.34 \\
SW     & 13.088 &  31.616 & 0.48 & 0.35 \\
S           &  13.095 & 31.587 & 0.56 & 0.41  \\
SE     & 13.103 &  31.823 & 0.37 & 0.27 \\
                    &           &                       &                      &                      \\
\multicolumn{2}{l}{Arp\,220} & & &   \\
W       & 57.094 & 31.876 & 30.42&   30.2 \\
E       & 57.091 &  32.030  & 13.38 &  13.3  \\
\hline
\end{tabular}
\end{center}
\scriptsize{Notes: (a) Positions of IC\,860 are relative to RA = 13$^h$15$^m$and Dec = $+$24\degr37\arcmin. 
Positions of IRAS\,15107+0724 are relative to RA = 15$^h$13$^m$ and Dec = $+$07\degr13\arcmin. 
Positions for Arp\,220 are relative to RA = 15$^h$34$^m$ and Dec = $+$23\degr30\arcmin.}
\label{tab:table1}
\end{table}

The observations for Arp\,220 were obtained in May 2004 and lasted for a total of 32 hrs with a target 
on-source time of 24.6 hrs. 
The flux density calibrator used was 3C\,286 (7.60 Jy,) the baseline calibrator was OQ\,208 (0.99 Jy), and the 
phase reference source were 1511+238 (0.80 Jy)  and  1530+239  (0.28 Jy).
The observations for Arp\,220 had an observing bandwidth of 16 MHz centred at a frequency of 4744.4 MHz giving 
a centre velocity of 5292.6 \kms close to the peak velocity in the single-dish spectrum of 
5430 \kmss. For Arp\,220 at a radial velocity 5375 \kms and distance of 73 Mpc, 
the conversion from angular size to projected linear size is 356 pc/arcsec. 
The projected separation between the two continuum peaks in Arp\,220 is 0.9 arcsec or 321 pc \citep{Sakamoto08}.

The data reduction procedure followed the standard AIPS routines for flagging, calibration,  bandpass calibration, 
self-calibration of phase calibrator, and imaging. 
The incremental phase and amplitude corrections obtained from the nearby phase calibrators were 
applied to the target source before imaging. 
The 31 channels in the observing bands had a channel width of 0.5 MHz, which corresponds to a velocity resolution 
of approximately 31.04 \kmss. 
The synthesized beam for IC\,860 is 0.081\arcsec $\times$ 0.058\arcsec 
(18 pc $\times$ 13 pc) and 0.101\arcsec $\times$ 0.071\arcsec (26 pc $\times$ 18 pc) for IRAS\,15107+0724.
For Arp\,220 the corresponding channel velocity resolution is  31.06 \kms and the synthesized beam size of Arp\,220 
is 0.079\arcsec $\times$ 0.068\arcsec (25 pc $\times$ 23 pc).

\begin{figure}
\begin{center}
\includegraphics[width=0.95\columnwidth]{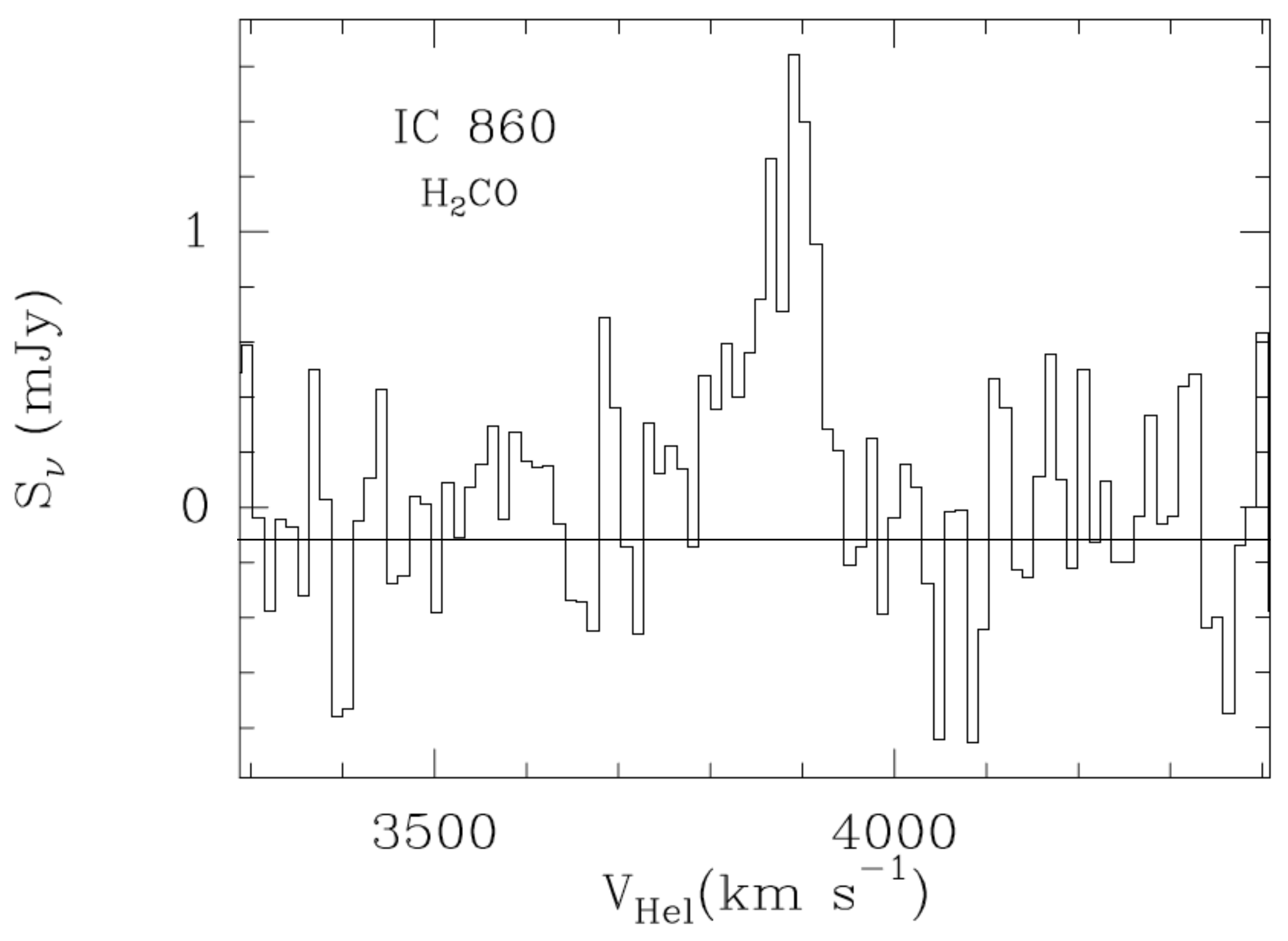}
\caption{A single-dish spectrum of the formaldehyde emission in IC\,860 obtained at 4.83 GHz with the Arecibo radio 
telescope in 1993 \citep{Araya04}. 
}
\label{fig:IC860sd}
\end{center}
\end{figure}

The continuum structure of the sources has initially been obtained from the bandpass-corrected data cube 
by averaging two or three channels at the edge of the cube that show no evidence of the remaining bandpass structure 
and no discernible line emission.
However, in order to obtain more accurate outer contours of the continuum structure, the centre 75\% of the channel maps 
have been used to represent the continuum structure for IC\,860 and  IRAS\,15107+0724, which then includes 
the weak line emission at the centre locations.
The peak of the line emission for Arp\,220, and also IC\,860 and IRAS\,15107+0724, is found to be close to 
the centre of the band and the line emission in the centre channels has been obtained by subtraction of a 
flat continuum structure based on the edge channels across the whole cube. A flat baseline subtraction does 
not remove any existing channel structure in the band and only brings the continuum level close to zero. 
Only for the final results for IC\,860 the continuum subtraction was based on only the low frequency edge channels 
in order to take into account the possible line emission near the high frequency edge.
For Arp\,220 the emission extends beyond the edge of the observing band, which may result in errors in the 
continuum subtraction. A larger observing band of these MERLIN observations would have helped to take away 
any uncertainty continuum subtractions.

The formaldehyde line emission features in all three sources are known to be very weak from single-dish observations 
\citep[e.g.,][]{Araya04} and special attention has been paid during the data reduction process in order to arrive 
at {bf reliable} results. Some of the data has also been reduced using Miriad, which produced the same results. 
Simple tests such as determining the emission structures by stacking the emission line channels also produced 
the same results as using moments.
There is no evidence that any of the observed weaker features away from the nuclear region result from 
imperfect data reduction. 
All emission features in the maps are indeed found to be superposed on the radio continuum 
contours of the sources. 

The results of the current studies show some differences between line profiles obtained from the interferometric data 
and those obtained with single-dish experiments. Some differences may result from residual errors in calibration and 
continuum subtraction using a limited number of line-free edge channels. In addition, the presence of diffuse 
absorption against the radio continuum may account for differences when sampled with different 
beam sizes. 

The analysis of the three target sources requires an evaluation of the observed brightness temperatures of the 
formaldehyde spectral components and the continuum structures. The brightness temperatures of a source 
component may be determined as \citep{KellermannOwen88}:
\begin{equation}
T_{\rm b} = 1.22 \times 10^{12} {S_{\nu}} (\theta_{\rm maj} \theta_{\rm min} \nu^{2})^{-1} (1+z)\, {\rm K}, 
\end{equation}
where the observing frequency $\nu$ is in unit of GHz, the component flux density $S_{\nu}$ has a unit of Jy, the 
source sizes $\theta_{\rm maj}$ and $\theta_{\rm min}$ are in milli-arcseconds. 

In this paper, distances were determined using a cosmological model with H$_0$ = 73 \kms Mpc$^{-1}$, 
$\Omega_{\rm M} = 0.27$ and $\Omega_\Lambda = 0.73$ \citep{Spergel07}.

\section{Formaldehyde in IC\,860 - IRAS 13126$+$2452}
\label{sec:IC860}

\subsection{IC\,860 - Formaldehyde Emission}

The 4.83~GHz formaldehyde K$_a$ = 1 emission in IC\,860 - IRAS 13126$+$2452 has first been observed with the 
Arecibo Observatory with a peak flux density of 2.0 - 2.2 mJy  \citep{Baan93}. 
A representative spectrum shows an asymmetric profile centred at 3860 \kms with a total width of 160 \kms as 
presented in Figure \ref{fig:IC860sd} \citep[from][]{Araya04}.   
IC\,860 also exhibits {\rm HI} absorption centreed at 3866 \kms as well as a combined OH spectrum with 
absorption at 3850 \kms and emission at 4000 \kms \citep{Schmelz86}. These spectra of IC\,860 cover 
a total velocity range from 3700 to 4100 \kmss.
Similarly, the formaldehyde spectrum of IC\,860 may also extend from 3550 to 4100 km/s with a weak emission 
feature at 3580 \kms and weak absorption features at 3700 and 4050 \kmss. This signature may also be recognised 
in the spectrum obtained with the Green Bank Telescope \cite{Mangum13}. 
These weak features may be further verified using single-dish or interferometric data.

\begin{figure}
\begin{center}
\includegraphics[width=0.9\columnwidth]{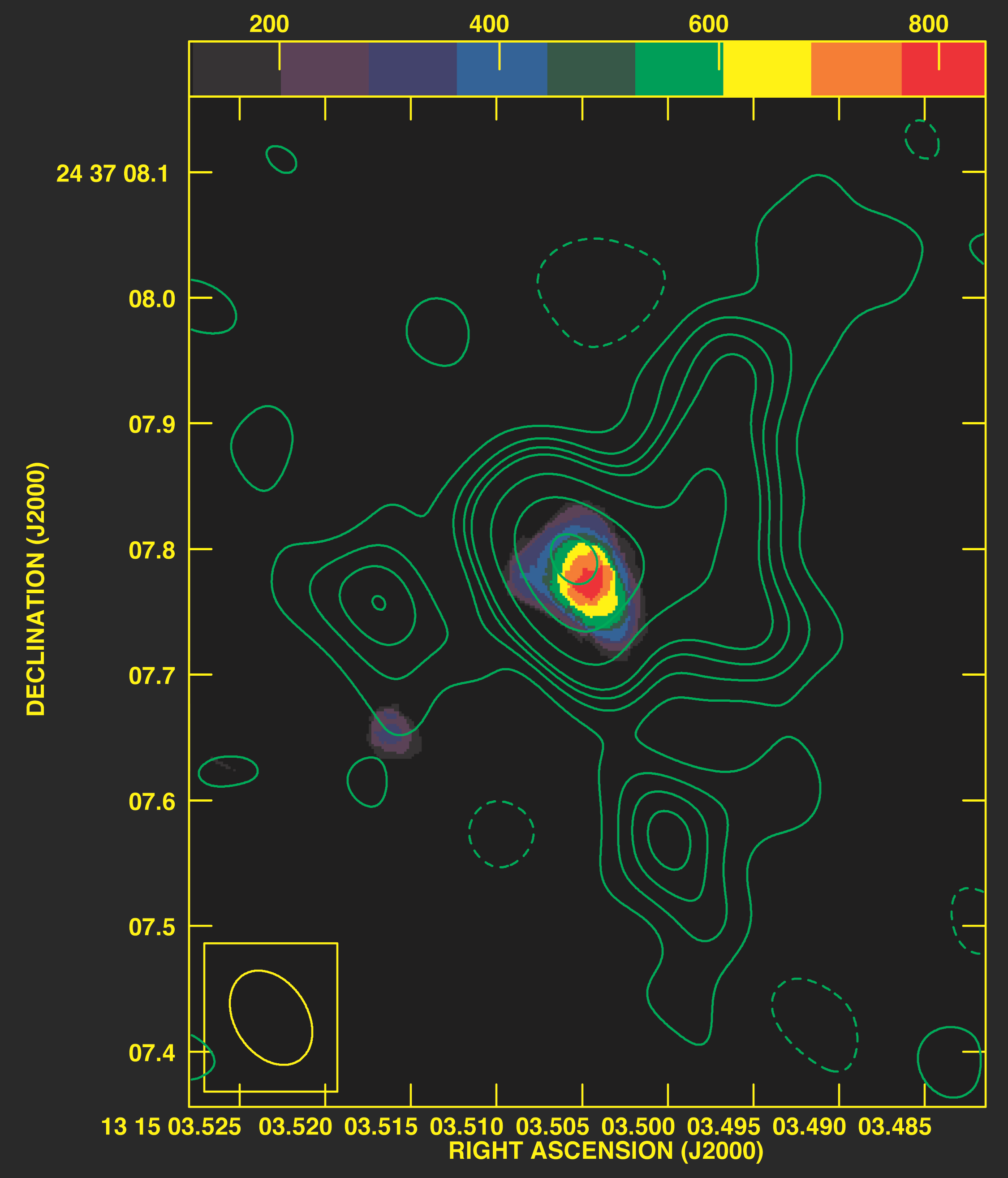}
\vspace{2mm}
\caption{The H$_2$CO emission structure superposed on the nuclear continuum structure of IC\,860. 
The contoured continuum emission structure has a peak flux density of 6.42 mJy beam$^{-1}$, which includes 
the weak line emission. The actual continuum peak intensity is 4.68 mJy beam$^{-1}$.
The continuum contour levels are 0.18 mJy beam$^{-1}$ $\times$ (-1, 1, 2, 3, 4, 8, 16, 32) with a rms noise 
in the map of 0.077 mJy beam$^{-1}$.
The moment 0 map of the H$_2$CO emission structure is presented as a colour scale map with a range of 
120$-$840 mJy beam$^{-1}$ \kmss. Two emissions components are found in the map as the prominent 
Centre region and the weaker SouthEast region. The integrated peak of the Centre line emission converts 
to 1.74 mJy beam$^{-1}$.}
\label{fig:IC860mom0}
\end{center}
\end{figure}
\begin{figure}
\begin{center}
\includegraphics[width=1\columnwidth]{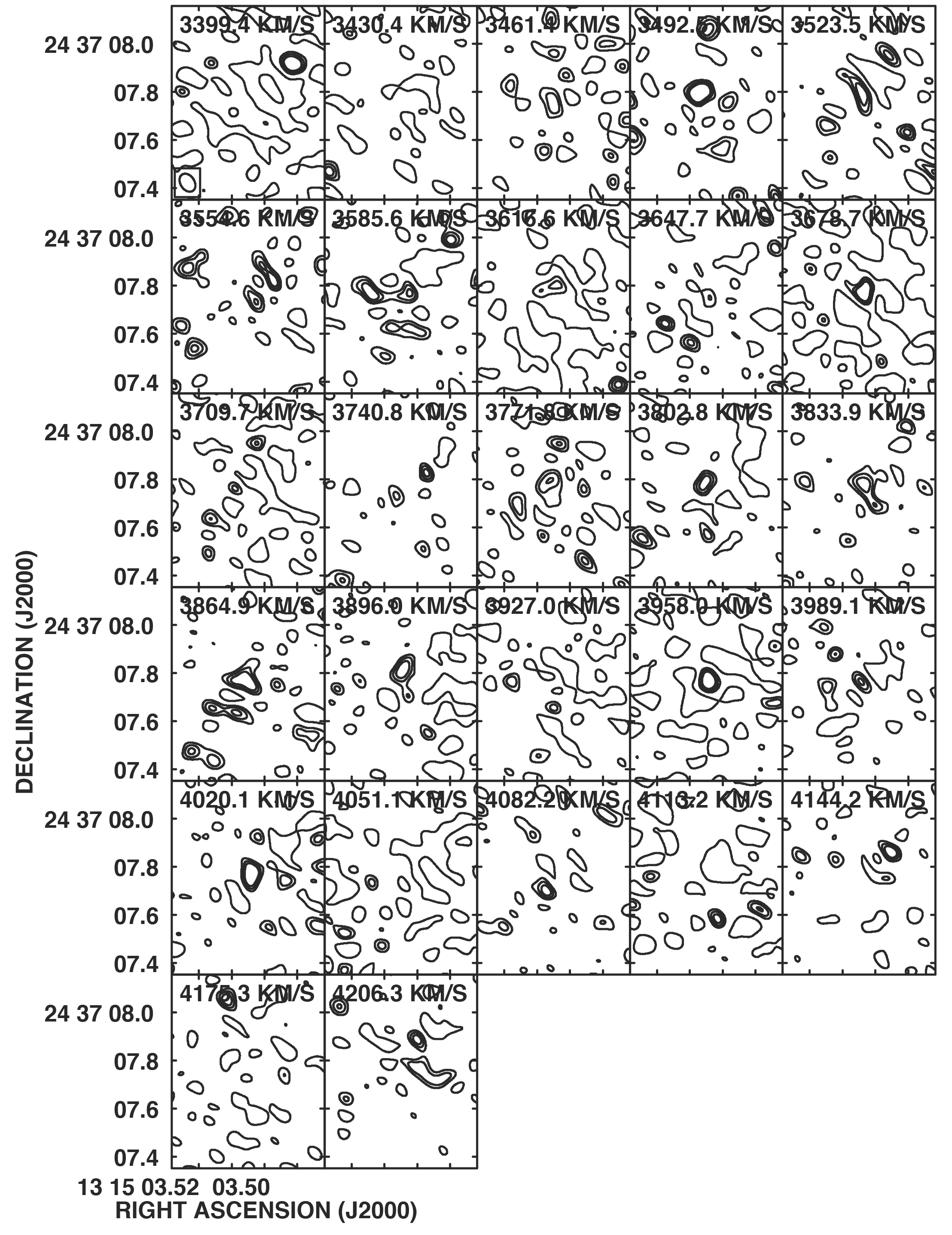}
\caption{Channel maps of IC\,860. The contour levels for the 27 central channels emphasise the negative (single 
solid line) contours as well as the strongly positive (multiple concentric) contours of the maser emission features. 
The contour levels are 0.5 mJy beam$^{-1}$ $\times$ (-1, 1, 2, 3) such that the extended single contours are negative.
}
\label{fig:IC860chan}
\end{center}
\end{figure}
\begin{figure}
\begin{center}
\includegraphics[width=0.9\columnwidth]{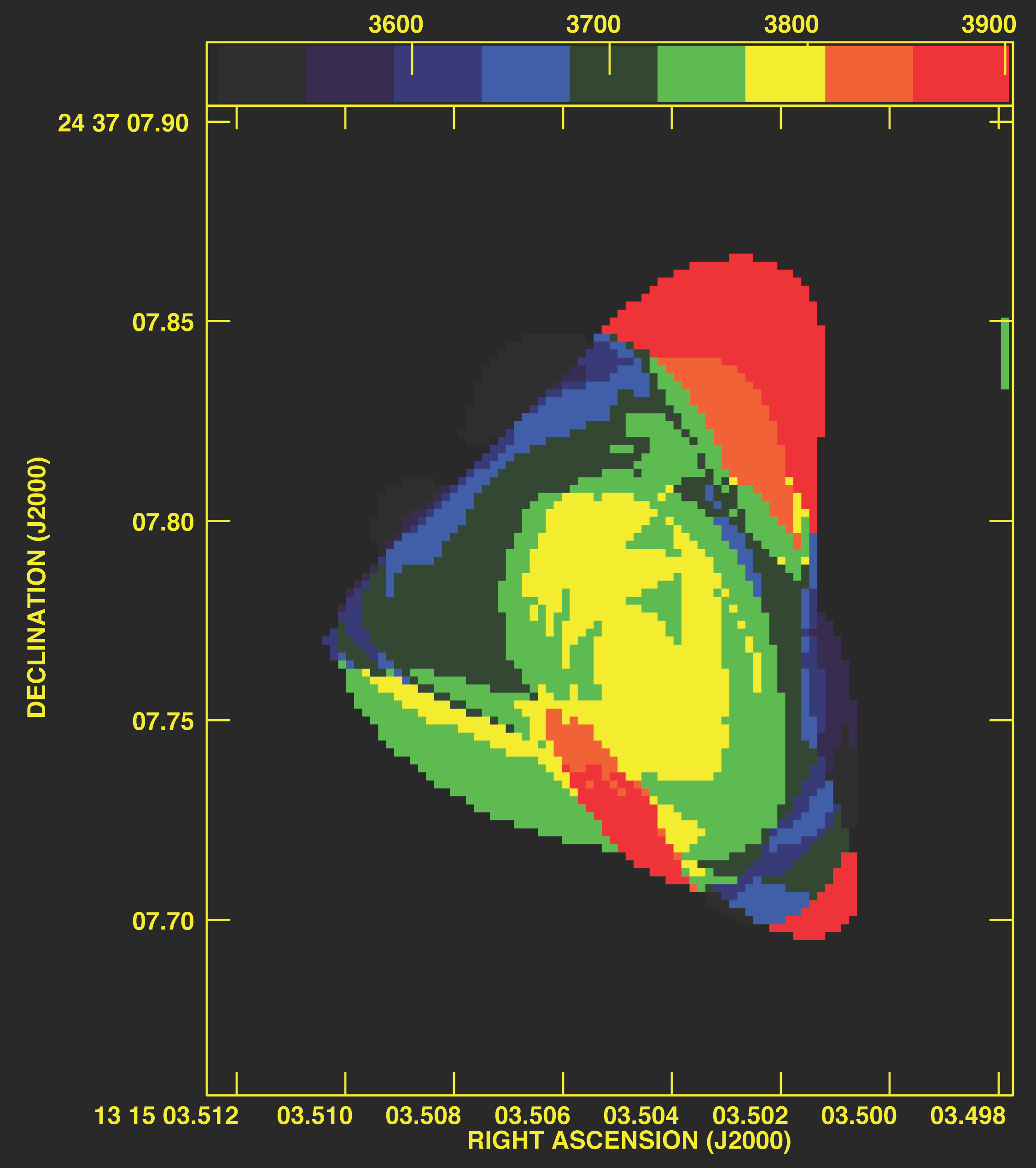}
\vspace{2mm}
\caption{The velocity structure of the central H$_2$CO emission structures of the nuclear region of IC\,860. 
The first moment map shows colour contours between 3500 and 3900 \kmss.  
The velocity structure may reveal a superposition of point sources but does show an overall velocity gradient.
}
\label{fig:IC860mom1}
\end{center}
\end{figure}
\begin{figure}
\begin{center}
\includegraphics[width=0.6\columnwidth,angle=-90]{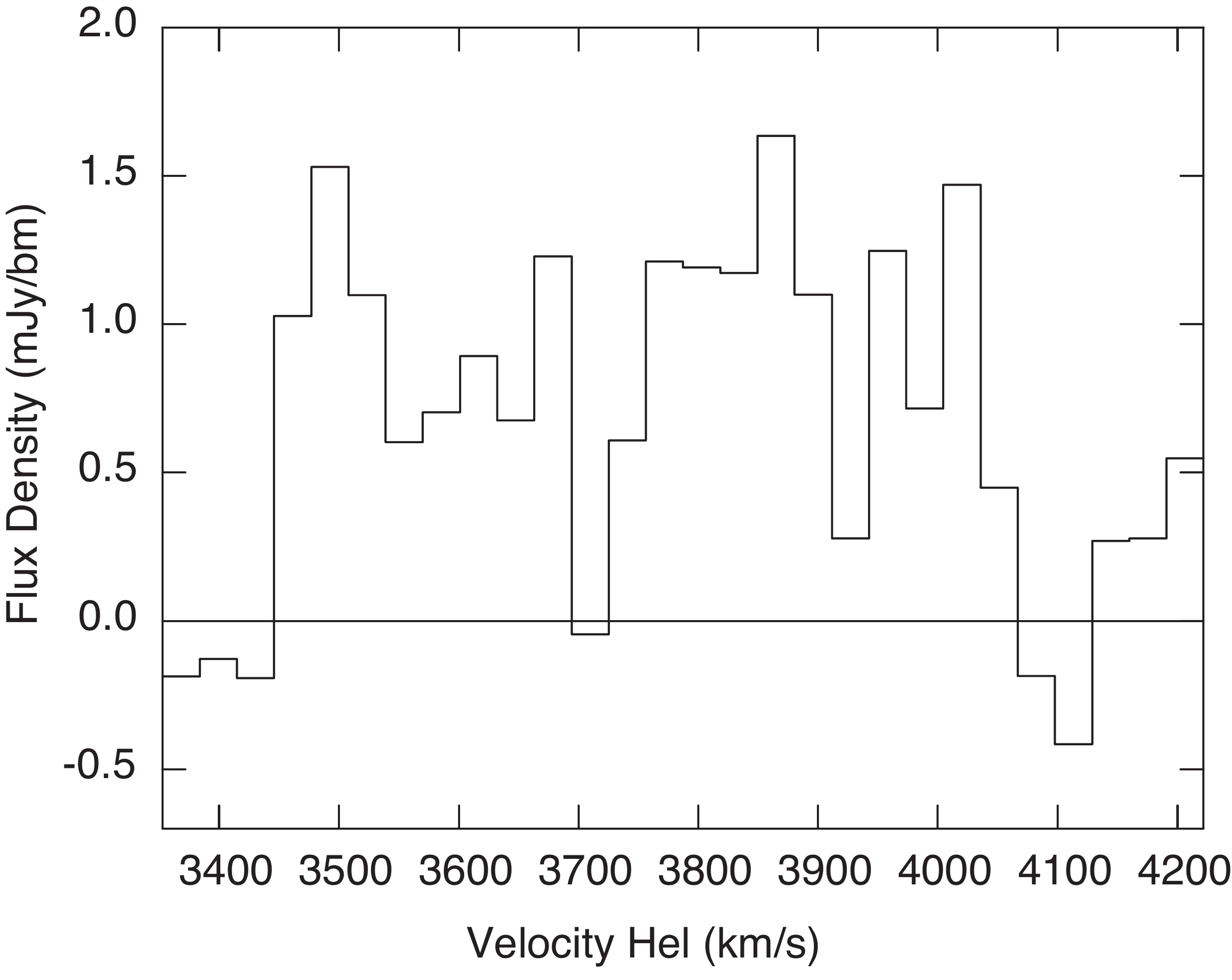}
\includegraphics[width=0.6\columnwidth,angle=-90]{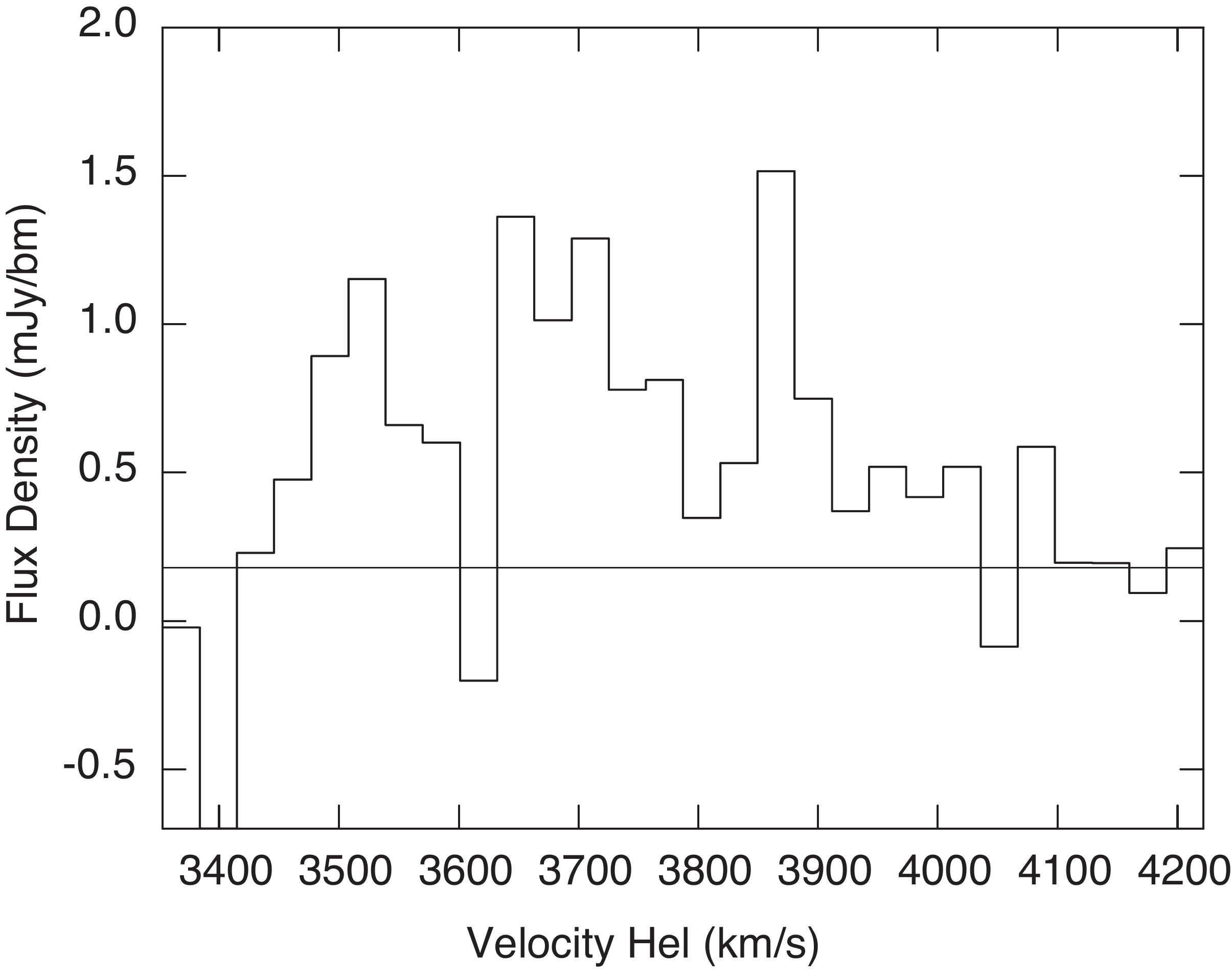}
\caption{The integrated spectra of the formaldehyde emission in IC\,860 integrated of the Centre emission region 
and the SouthEast region.  
(top) The Centre region exhibits three distinct velocity components at V = 3590,  3830, and 3990 \kmss. 
The 3590 \kms component has a weak counterpart in the single-dish spectrum of Fig. \ref{fig:IC860sd}, while the other 
components find their counterparts in the main feature of the single-dish spectrum and a possible high velocity wing.
(bottom) The spectrum at the SouthEast region shows three distinct components at shifted velocities of V = 3520, 3720 and 3870 \kmss.}
\label{fig:IC860spec-c}
\end{center}
\end{figure}

The formaldehyde line emission structure of IC\,860 is presented in Figure \ref{fig:IC860mom0} as a zeroth-moment 
colour map overlaid on the radio continuum map. The moment map incorporates line emission features above a 
threshold of 0.8 mJy beam$^{-1}$, while the rms in the line channel maps (away from the location of the source) is 
0.31 mJy beam$^{-1}$.  
The central line emission region extends 31 pc at PA = 130\degr{} with an additional extension to the east at PA = 270\degr{}. 
The region is centred on the nuclear continuum peak and its orientation agrees well with that
of the 2Mass image of the galaxy of PA = 150\degr{} \citep{Skrutskie06} and lies perpendicular to the northwest continuum 
extension.
One weaker and significant emission feature, SouthEast of the nucleus, is found in the zeroth-moment map inside the 
continuum contours. 

The underlying continuum structure of the nuclear region of IC\,860 consists of a compact component and
extended components towards the northwest, the southwest, and to the east. 
With an integrated flux density 15.31 mJy, the compact component has a peak flux 
density of 4.68 mJy beam$^{-1}$. The brightness temperatures, $T_{\rm b}$, of the central component ($5.13 \times 
10^4$ K) and the other components (in the range 0.45 to 0.78 $\times 10^4$) K are consistent 
with the occurrence of star formation in all continuum components \citep{Condon92}.
In order to better define the outer continuum structure, the (weak) line emission close to the nuclear centre has 
been included in the continuum map in Figure \ref{fig:IC860mom0}, which results in an increased peak flux density of 6.42 mJy beam$^{-1}$.
The parameters for the nuclear and circumnuclear components are presented in Table \ref{table1}.

\subsection{IC\,860 - Formaldehyde Emission Structure}

The channel maps of the line emission for IC\,860 have been presented in Figure \ref{fig:IC860chan}.
The negative (single solid line) contours indicate the general noise level of each of the maps as well as 
the location of (extended) structures possibly representing weak absorption against the radio continuum of the source. 
 The features with multiple positive contours identify the emission features located within the contours of the radio 
 continuum.  These feature have the approximate size of the resolving beam and represent very localised 
 emission regions with single or multiple maser sources that often fill a single spectral channel in the maps. 
Features above the second positive contour have been added to the Moment 0 map in Figure \ref{fig:IC860mom0}.

The first moment map of the central emission components in IC\,860 does not show a distinct velocity gradient 
(Fig. \ref{fig:IC860mom1}). 
Instead the colour contours indicate a dominant emission component with a velocity in the range 3730-3820 
\kms superposed on some larger scale background velocity structure with 3500 \kms at the East and West 
edges and 3850 \kms in the North and South. 

\subsection{IC\,860 - Spectral Characteristics}

The integrated spectral profile of the H$_2$CO line emission in the nuclear Centre region of IC\,860 is presented in 
Figure \ref{fig:IC860spec-c}(a) and shows three distinct emission components. 
The features centred at 3830 and 3990 \kms (ranging from 3720 - 4050 \kmss) are in rough 
agreement with the single-dish emission components peaking at 3890 \kms in Figure \ref{fig:IC860sd}, although the 
high-velocity edge is less prominent in the single-dish spectrum. 
The broad feature at 3590 \kms with a mean flux of 0.95 mJy beam$^{-1}$ (line width of 230 \kmss) only 
has a weak counterpart in the single-dish spectrum.  
The integrated spectrum of the SouthEast region is displayed in Figure \ref{fig:IC860spec-c}(b).  
This spectrum shows emission peaks at 3700 and 3880 \kms but also the smaller feature at 3580 \kms that is also 
found in the Centre region.

In addition to the main emission features seen in the single-dish spectrum of Figure \ref{fig:IC860sd}, the 
emission component close to 3580 \kms does not have a strong counterpart in the single-dish spectrum 
and occurs precariously close to the edge of the observing band. 
Careful inspection of the data reduction procedure shows that this emission is not the result the bandpass 
calibration or the flat continuum subtraction using the edge channels. (A great effort has been done to make this 
feature disappear but it would not.)
While the strength of this  emission component remains unexpected, we suggest that the feature really exists and 
is compensated in the single-dish spectrum by absorption in the nuclear region in certain channels. 

The systemic velocity of IC\,860 is nominally at 3347 \kmss \citep{Haynes97}, but all known OH and HI 
emission and absorption features are found close to a higher velocity of 3911 \kmss. 
While the 2Mass image does not show clear signs of a galaxy interaction \citep{Skrutskie06}, these two velocities 
may indicate the interactive nature of the IC\,860 system consistent with its large FIR luminosity. 
The observed velocity component at 3580 \kms would be consistent with the maser emissions occurring closer 
to the systemic velocity.

Close to the peak of the continuum emission in Figure \ref{fig:IC860mom0}, the estimated amplifying optical 
depths for the spectral features vary as $\tau = 0.34 - 0.46$ and the brightness temperatures range from 
3.0 to 5.2 $\times 10^4$ K. The spectral components in the weaker SouthEast region show  an optical 
depth ranging from 1.11 to 1.90 with brightness temperatures ranging from 2.2 to 4.2 $\times 10^4$ K (Table \ref{table2}). 
The regions with the highest flux densities and the lowest gains are located towards the peak of the continuum 
distribution, which is consistent with a stronger background requiring a lower amplifying gain.

The distinct components at the centre and SouthEast locations between 3650 and 4000 \kms contribute to emission in the 
integrated single-dish spectrum (Fig. \ref{fig:IC860sd}).  
A superposition of the emission (and possible absorption) across the face of the nuclear continuum could
account for the differences between the single-dish spectrum and the current MERLIN spectra.  
However, the reality of the weak emission feature at 3580 \kms in the single-dish spectrum and the 
feature found in the current data does require further verification with more sensitive observations 
having a larger observing bandwidth.

\begin{figure}
\begin{center}
\includegraphics[width=1\columnwidth]{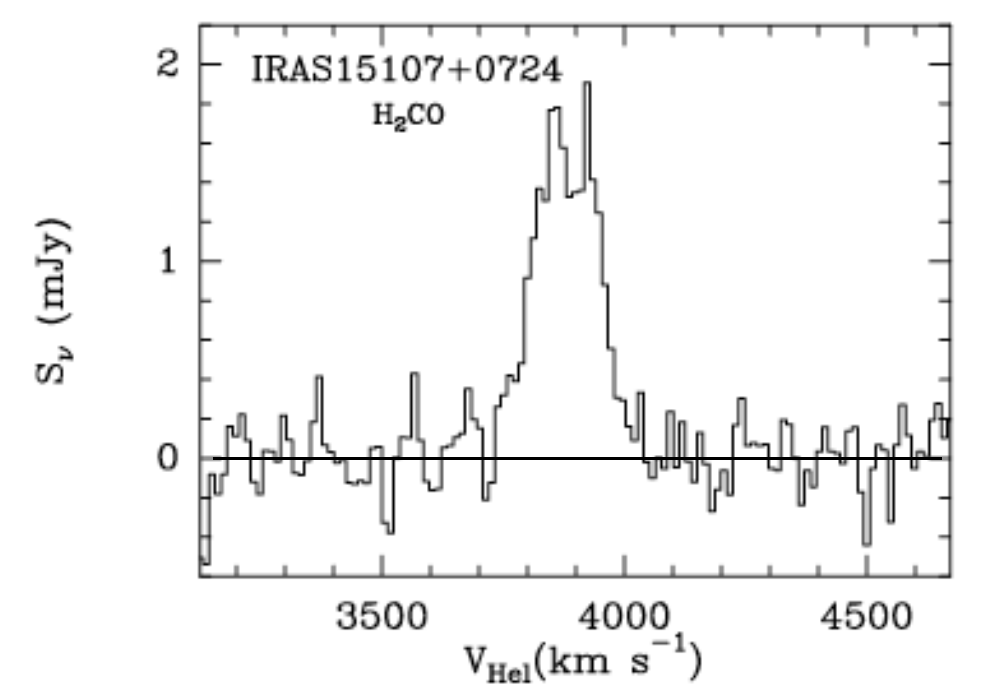}
\caption{A single-dish spectrum of IRAS\,15107+0724 obtained with the Arecibo radio telescope.  
Taken from \citep{Araya04}}
\label{fig:IR15107spec}
\end{center}
\end{figure}

\section{ Formaldehyde in IRAS\,15107$+$0724}
\label{sec:15107}
 
\subsection{IRAS\,15107 - Formaldehyde Emission}

IRAS\,15107+0724 has been first detected at Arecibo Observatory \citep{Baan93} with a peak flux of 1.91 mJy. 
A representative spectrum, presented in Figure \ref{fig:IR15107spec} \citep{Araya04}, shows a spectral line with
a mean velocity of 3880 \kms and a total width of 320 \kmss, but clearly consisting of two components. 
No other tentative features can be seen in the spectrum. 
An OH spectrum also shows two emission components centred at 3782 and 3900 \kmss, 
whereas the HI spectrum shows a broad absorption feature centred at 3902 \kms  \citep{Baan87}. 
The galaxy itself has a systemic velocity of 3897 \kmss.

\begin{figure}
\begin{center}
\includegraphics[width=0.90\columnwidth,angle=-90]{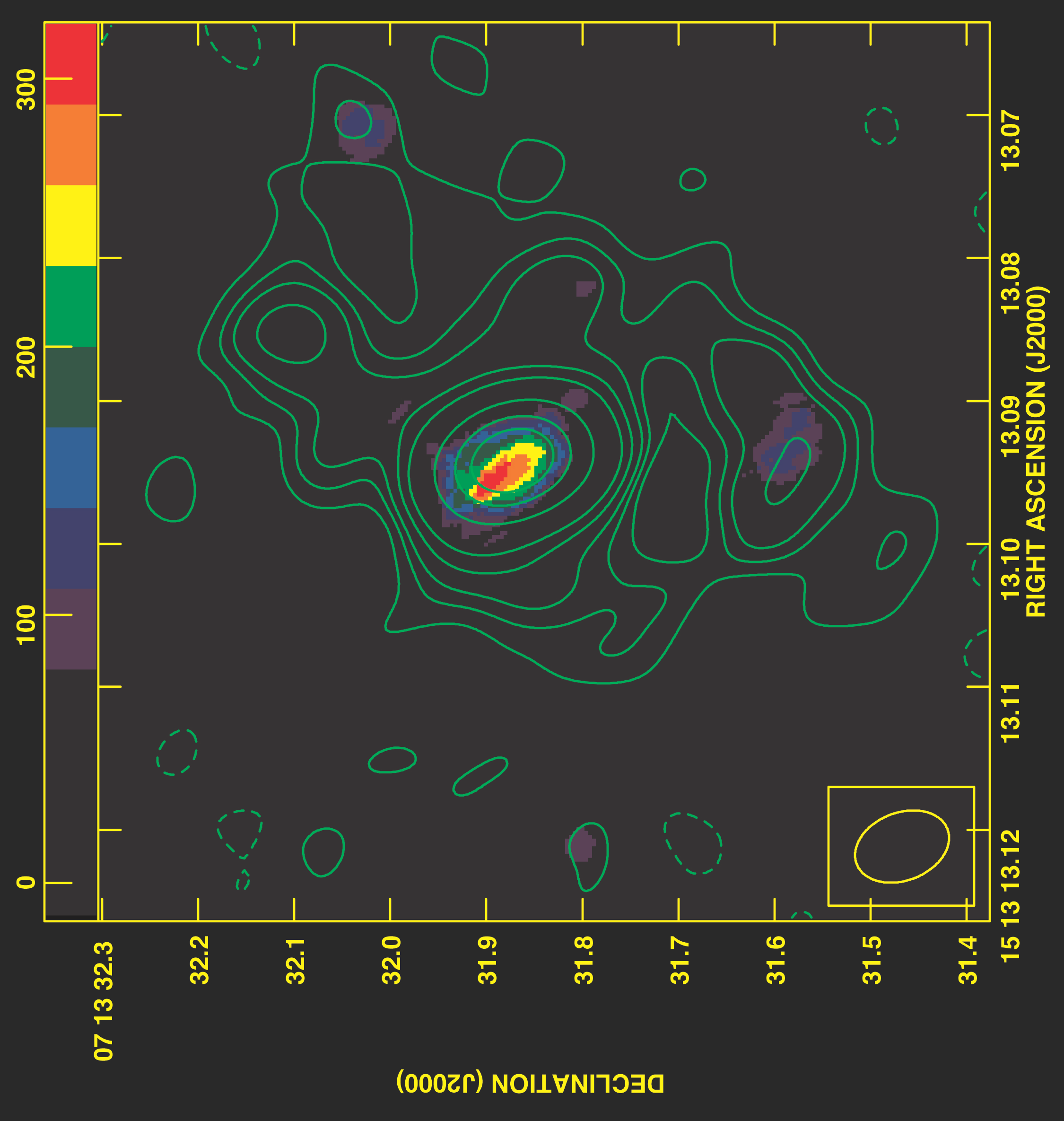}
\caption{The formaldehyde line emission and radio continuum structure of the IR\,15107+0724. 
A continuum source shows a central component with a peak flux of 7.92 mJy/beam surrounded by 
distributed emission regions resembling an ring or inner spiral arms. The contour levels are 0.11 mJy/beam 
$\times$ (-1, 1, 2, 3, 5, 8, 24, 48, 72).  The rms in the continuum map is 0.051 mJy/beam.
The zeroth moment of the formaldehyde emission is presented as a superposed colour scale map. 
The peak integrated flux is 308.8 mJy*km/s and the colour scale runs from 60 to 320 mJy*km/s. 
Three distinct emission regions are found within the continuum contours at the Centre position and at the 
South and NorthWest positions.}
\label{fig:15107comb}
\end{center}
\end{figure}
\begin{figure}
\begin{center}
\includegraphics[width=1\columnwidth]{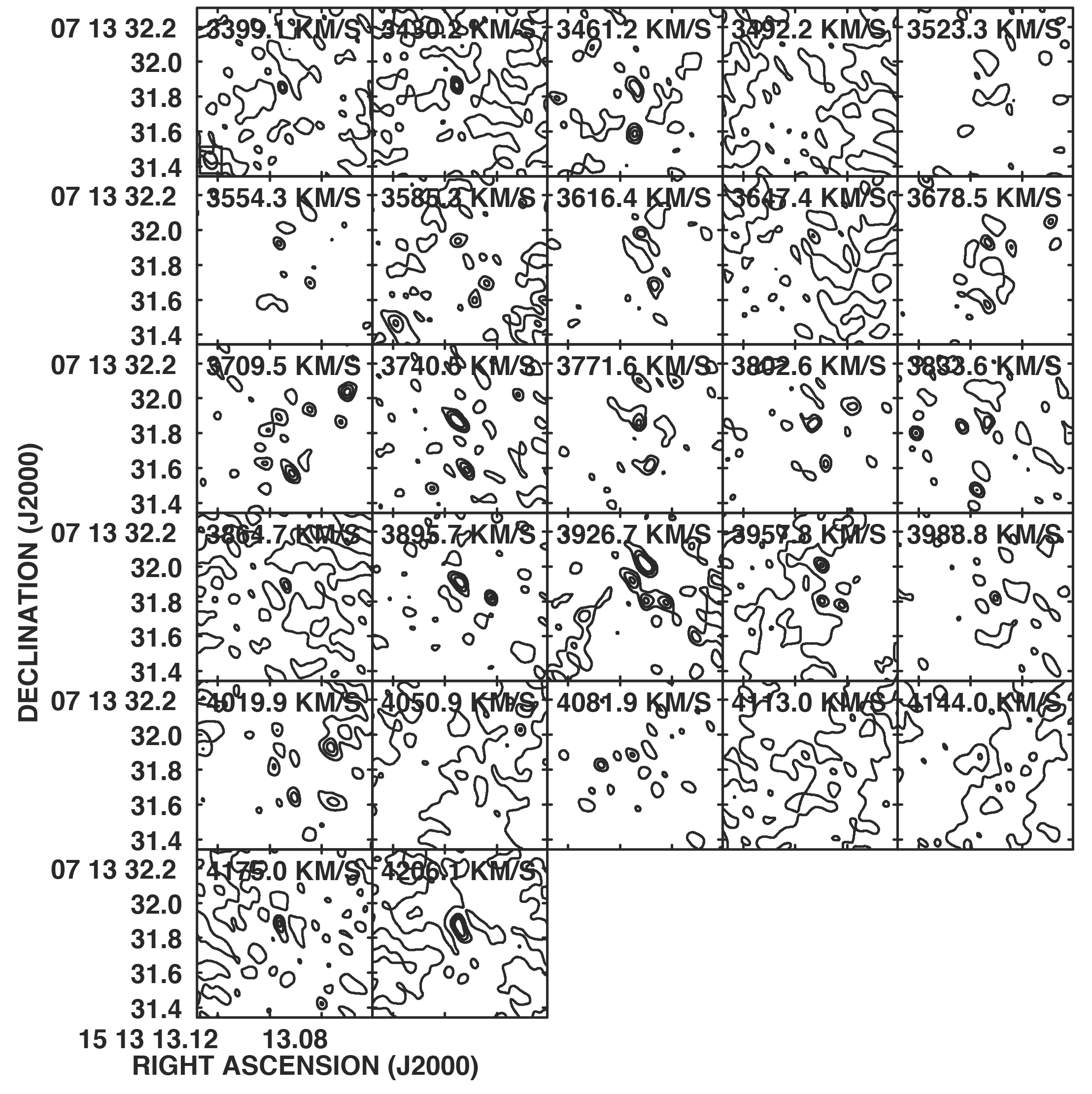}
\caption{Channel maps of IR\,15107$+$0724. The contour levels for the 27 central channels 
emphasize the negative (single solid line) contours as well as the strongly positive (multiple concentric) 
contours of the maser emission features. 
The contour levels are 0.2 mJy beam$^{-1}$ $\times$ (-2, 4, 6, 8).}
\label{fig:IR15107chan}
\end{center}
\end{figure}

\begin{figure}
\begin{center}
\includegraphics[width=0.85\columnwidth,angle=0]{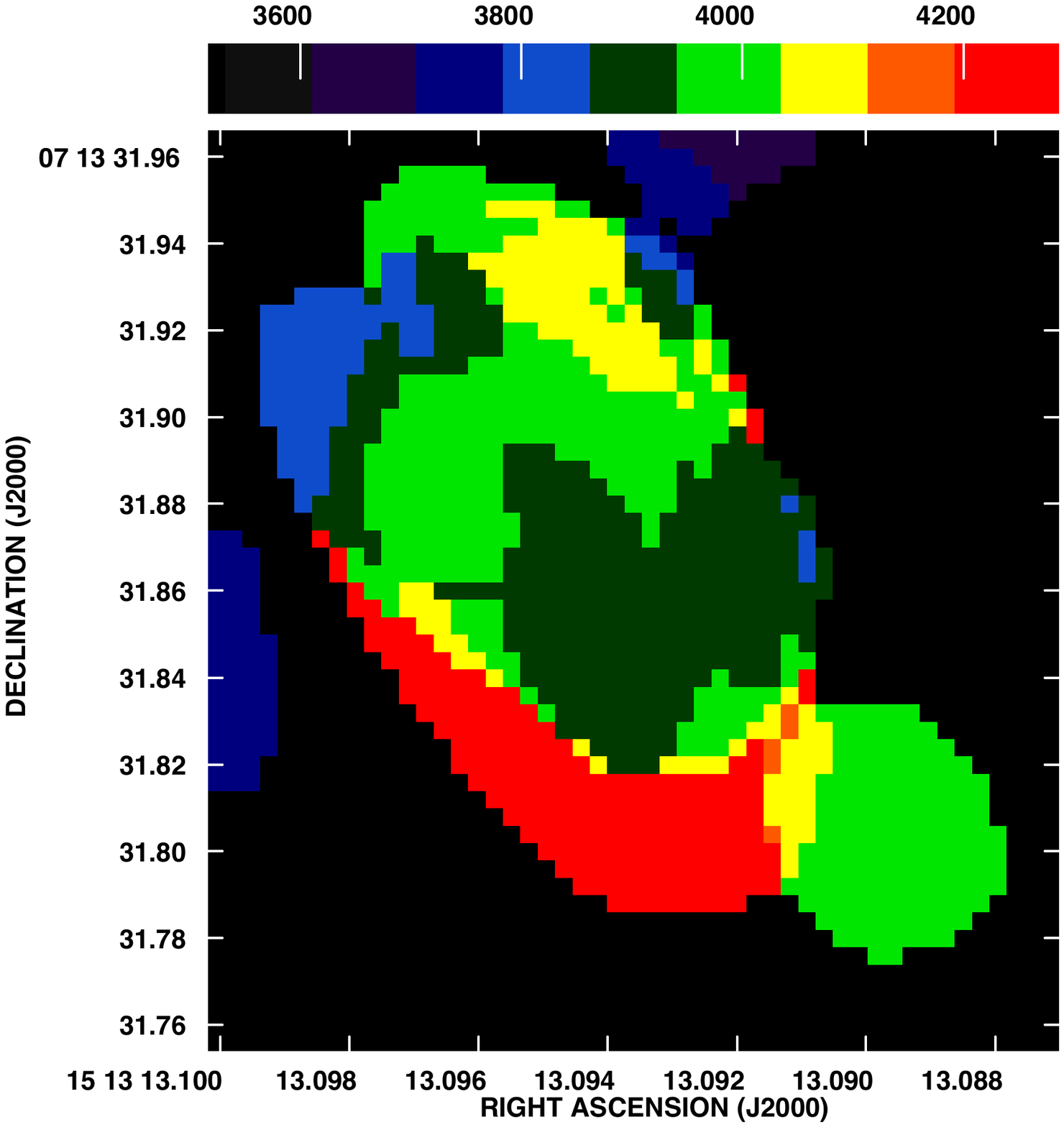}
\caption{The velocity field of the formaldehyde emission region at the Centre of IRAS\,15107+0724. 
The velocity colours in the first moment map range between 3750 and 4050 \kmss. 
No clear organised motion can be seen across the Centre emission region except for a (possible) velocity gradient 
covering 250 \kms from Northeast to Southwest. 
A dominant central component exists close to the systemic velocity of 3850 \kms located close to the 
central continuum peak of the source. The emission and the associated velocity field seems dominated by a 
superposition of strong emission components. 
}
\label{fig:15107M1}
\end{center}
\end{figure}

\begin{figure}
\begin{center}
\includegraphics[width=0.6\columnwidth,angle=-90]{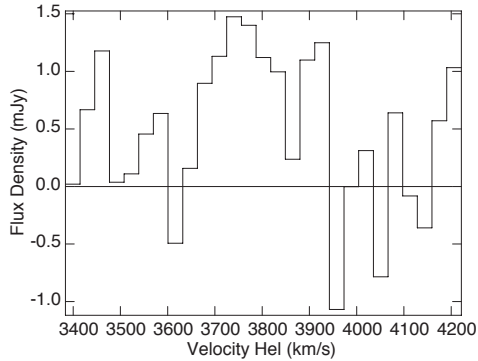}
\caption{Integrated formaldehyde emission spectrum at the Centre position of IRAS\,15107+0724.}
\label{fig:15107specC}
\end{center}
\end{figure}

The formaldehyde line emission in IRAS\,15107+0724 is presented as a zeroth moment colour overlay on 
the continuum contours in Figure \ref{fig:15107comb}. 
The line channels in the data cube have a rms noise of 0.17 mJy beam$^{-1}$ and the moment maps incorporate 
signals above a flux threshold of 0.4 mJy beam$^{-1}$.  
The main emission feature straddles the central continuum peak and has an extent of 52 pc at PA = 140\degr. 
This orientation agrees well with the orientation of the 2Mass image of the galaxy at PA = 155\degr \citep{Skrutskie06}.
Two other compact emission regions, South and NorthWest, are superposed on the extended radio continuum 
structure.  

\begin{figure}
\begin{center}
\includegraphics[width=0.6\columnwidth,angle=-90]{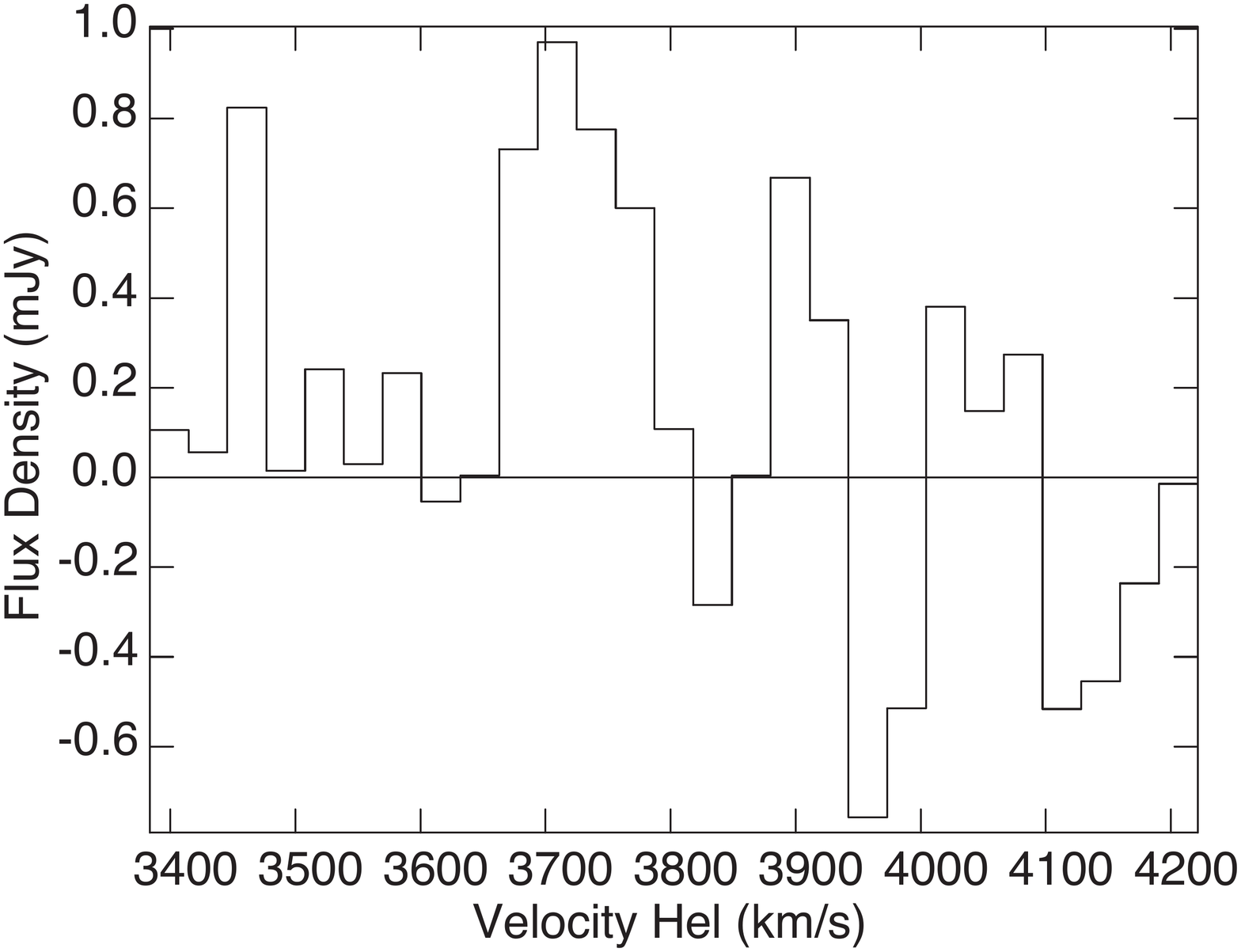}
\includegraphics[width=0.6\columnwidth,angle=-90]{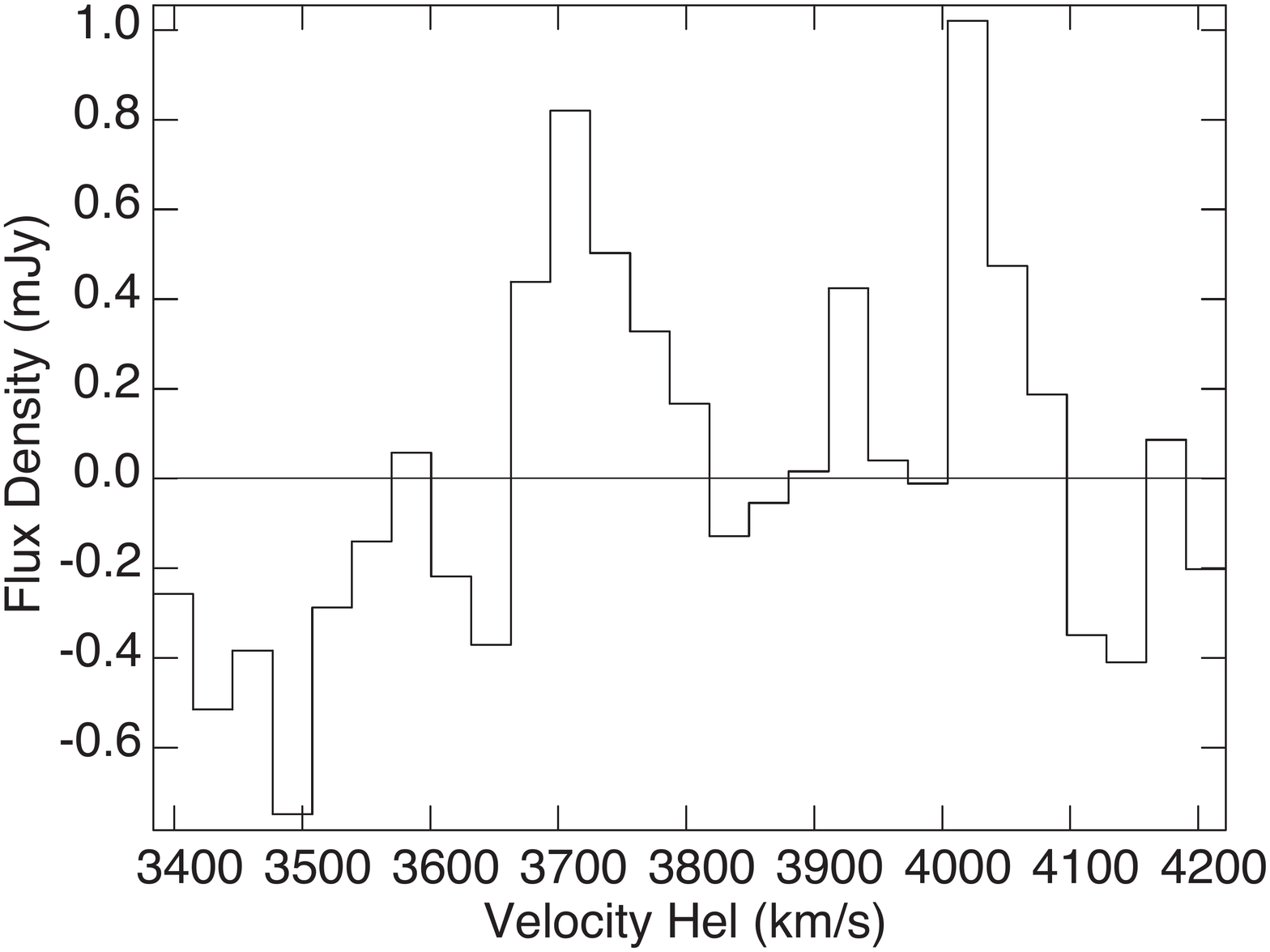}
\caption{Integrated formaldehyde spectra at the two additional emission components of IRAS\,15107+0724 
at the Southern component (top), and the NorthWest component (bottom). 
The spectra suggest that there is also absorption in the source and that individual emission components exist within the 
observed range of 3650$-$4050 \kms in the single-dish spectra (see Fig.\ref{fig:IR15107spec}).}.
\label{fig:15107spec3}
\end{center}
\end{figure}

The underlying continuum structure of IRAS\,15107$+$0724 shows a central nuclear component surrounded by a ring 
of emission tracing two inner spiral arms and resembling an S-shaped structure (Fig. \ref{fig:15107comb}). 
The total flux density of the source is 29.72 mJy with a peak of 12.57 mJy beam$^{-1}$. 
The brightness temperature at the nuclear source is $9.4 \times 10^4$ K, while the peaks in the surrounding 
structure have brightness temperatures in the range $(0.27$-$0.53)  \times 10^4$ K. 
These temperatures and the emission structure support the hypothesis that this entire S-shaped continuum structure 
traces star formation activity, although the presence of an embedded radio AGN at the nucleus cannot 
be excluded. 
In order to better define the outer continuum structure, the contour map presented in Figure 
\ref{fig:15107comb} also includes the central line emission.
The positions, the continuum flux densities, and the brightness temperatures of these identifiable components 
are presented in Table \ref{table1}.

\subsection{IRAS\,15107 - Formaldehyde Emission Structure}

The channel maps of the line emission for IR15107$+$0724 have been presented in Figure \ref{fig:IR15107chan}.
The negative (single solid line) contours indicate the general noise level of each of the maps as well as 
the location of possible absorption features against the radio continuum of the source. 
The features with multiple positive contours identify the localised emission features in the maps lying mostly 
 within the contours of the radio continuum.  
The features above the first positive contour have been presented in the Moment 0 map in Figure \ref{fig:15107comb}.
The emission structure of formaldehyde in IR\,15107$+$0724 is also made up of multiple (single) features that 
often fill only one spectral channel and are found across the centre of the velocity range. 
In addition, the presence of extended negative contours may be consistent with the concept that certain emission 
features are compensated by regions of absorption.

\subsection{IRAS\,15107 - Spectral Characteristics}

The velocity field in the first moment map shows a semi-organized NE-SW velocity 
gradient (PA = 145\degr) across the nuclear emission region (Fig. \ref{fig:15107M1}) with a central dominant 
component at 3850 \kms superposed close to the centre. 

The spectral signature of the central emission component in the zeroth moment map in Fig. \ref{fig:15107comb} has 
been presented in Figure \ref{fig:15107specC}. 
The central component contributes most to the integrated emission close to the systemic velocity with a peak at 3750 
\kmss, which is below the observed 3880 \kms velocity of the emission peak and the 3897 \kms systemic velocity.
The spectra for the two weaker emission components, South and NorthWest, are presented in Figure 
\ref{fig:15107spec3}.
Both the South and NorthWest emission regions show additional velocity components below (3710 \kmss) and above 
(4015 \kmss) the systemic velocity of 3820 \kmss.
These coarse velocity designations are quite consistent with the velocities  found at the Centre region, although 
the multiple velocity systems are reminiscent of a merger scenario.

\begin{figure}
\begin{center}
\includegraphics[width=1.0\columnwidth]{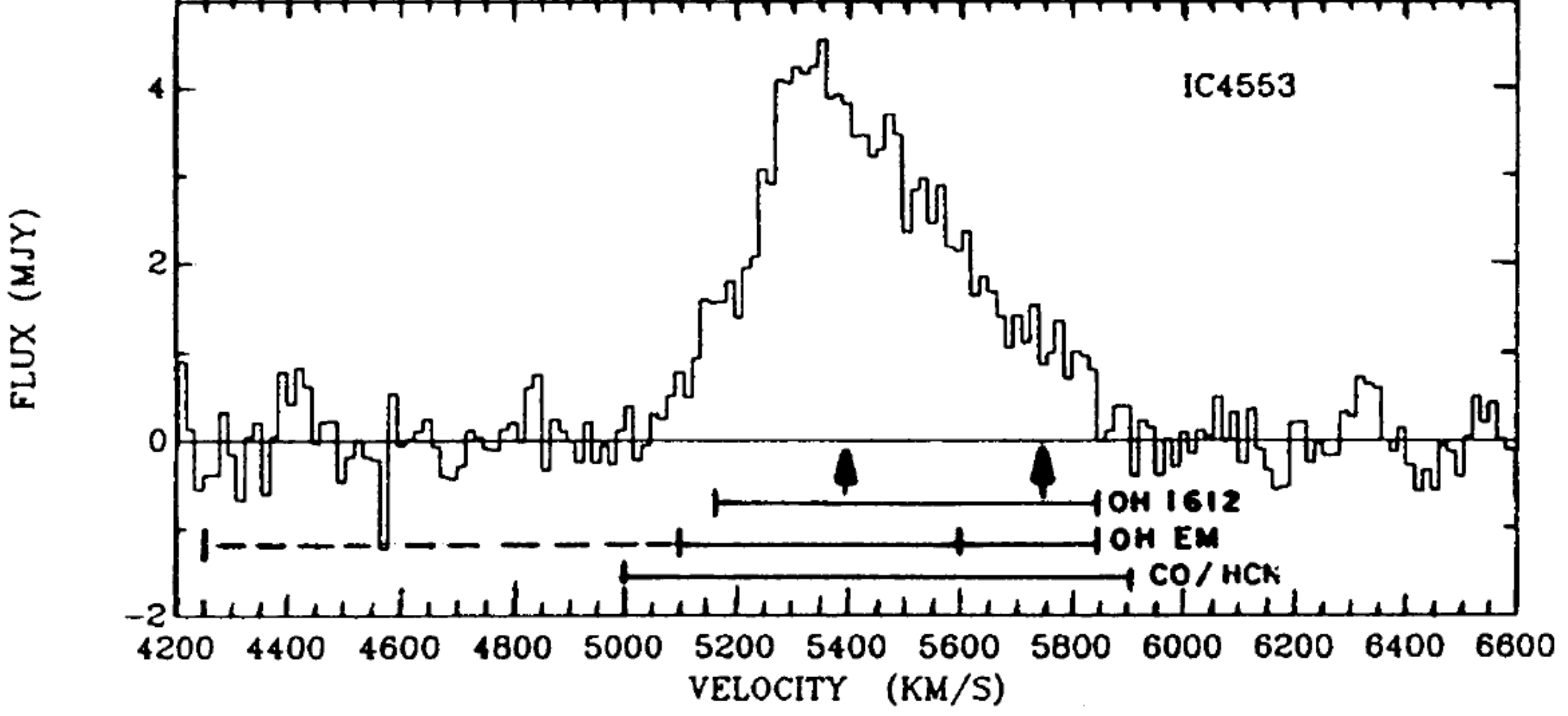}
\caption{A single-dish spectrum of Arp\,220 obtained with the Arecibo radio telescope. 
The two arrows indicate the systemic velocities of the West and East nuclei. The horizontal bars below the spectrum indicate the observed velocity range of the OH 1612 MHz absorption, the 
OH 1667/1665 MHz MM emissions, and the thermal CO(1-0) and HCN(1-0) emissions in the nuclear region. 
The dashed line indicates the range of the observed blueshifted OH outflow emission. 
}
\label{fig:ARarp220}
\end{center}
\end{figure}

\begin{figure*}
\begin{center}
\includegraphics[width=1\columnwidth,angle=-90]{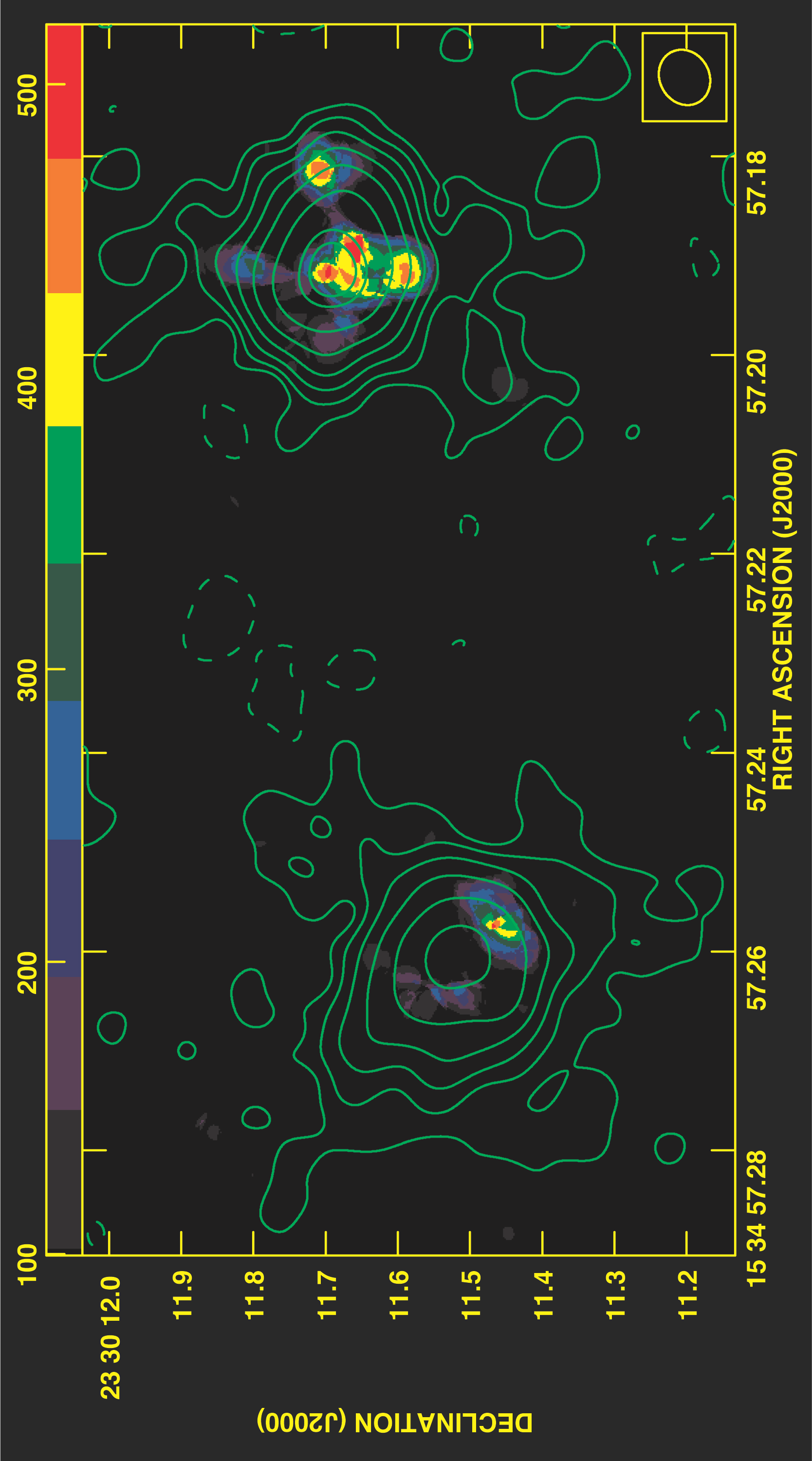}
\caption{A composite map of the continuum emission and the formaldehyde line emission in Arp\,220. 
The continuum structure is depicted with contours at levels 0.30 $\times$ (-1, 1, 2, 4, 8, 16, 32, 64, 80) 
mJy beam$^{-1}$ and a peak flux of 30.42 and 13.38 mJy beam$^{-1}$ for the western and eastern nucleus, respectively.
The colour scale zeroth moment image depicts a range of  0.1 to 0.56 mJy beam$^{-1}$ km s$^{-1}$. 
The line emission image show four emission regions at the Western nucleus: W-Centre, W-North, W-West, and W-South.
The Eastern nucleus displays two line emission regions: E-West and E-East.
}
\label{fig:A220comp}
\end{center}
\end{figure*}
\begin{figure*}
\begin{center}
\includegraphics[width=2.0\columnwidth,angle=0]{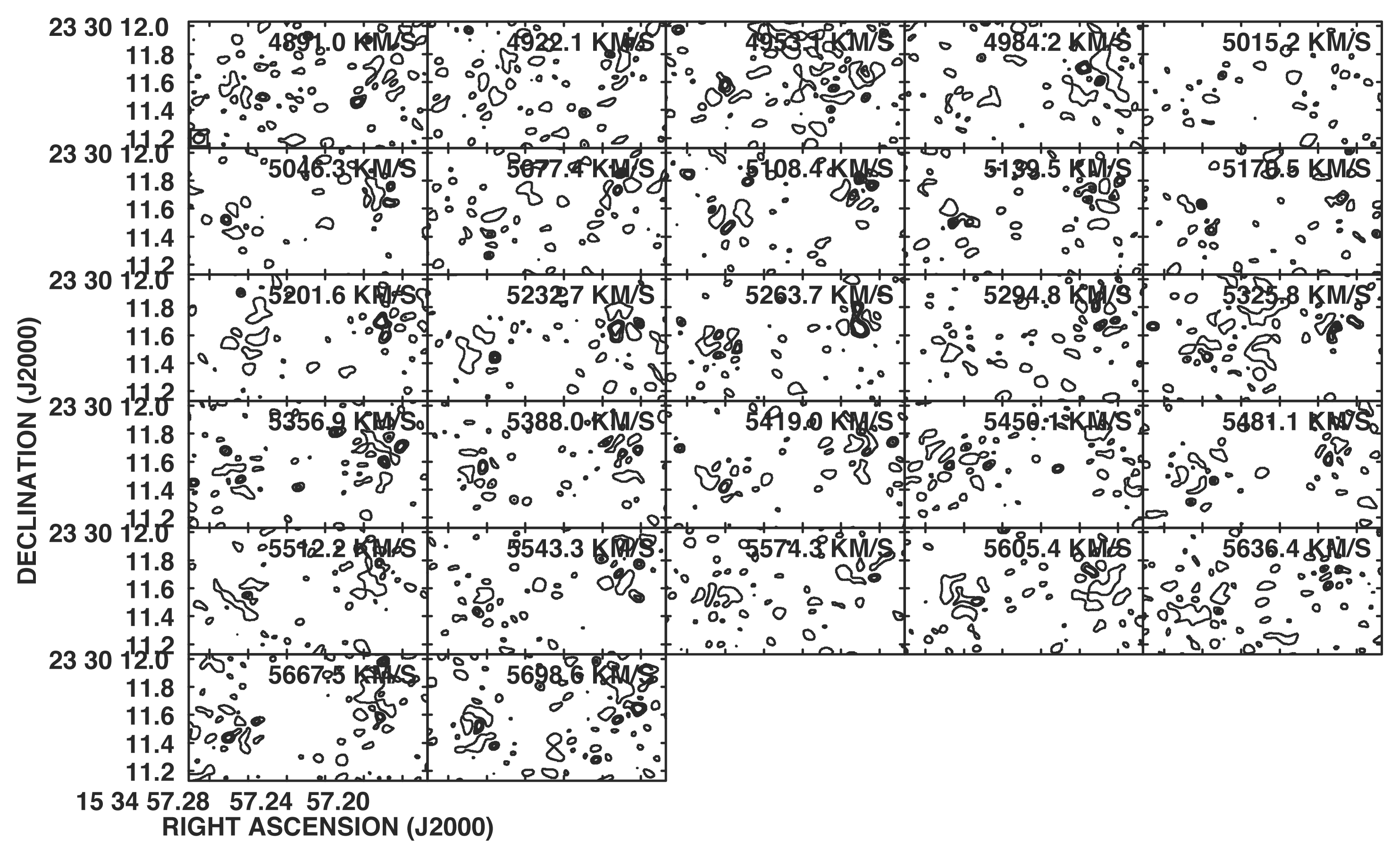}
\caption{Channel maps of Arp\,220. The contour levels for the 27 central channels 
indicate the negative (single solid line) contours as well as the strongly positive (multiple concentric) 
contours of the maser emission features. 
The contour levels are 0.3 mJy beam$^{-1}$ $\times$ (-2, 4, 6, 8) }
\label{fig:A220chan}
\end{center}
\end{figure*}

The characteristics of all emission regions within the continuum confines of IRAS\,15107$+$0724 have been presented 
in Table \ref{table2}. 
The peak line flux density at the nucleus is 1.45 mJy beam$^{-1}$ and the integrated fluxes at the three other regions 
are in the range 0.45 - 1.0 mJy beam$^{-1}$. 
The central line emission regions in IRAS\,15107$+$0724 superposed on the continuum peak have a brightness 
temperature in the range of 3.4 to 4.2 $\times 10^4$ K and apparent optical depths of only 0.16 to 0.20 
(see Table \ref{table2}).
The line emission features in the two other features have brightness temperatures higher than those of the radio 
continuum ranging from 0.6 to 4.0 $\times$10$^4$ K and they have higher optical depths ranging from 0.44 to 1.71. 
Although some larger optical depths are found, both the brightness temperatures and the deduced optical depths are 
consistent with modelling results with radiative pumping schemes (details in Sect. \ref{sec:pump}). 

The spectral features emission found in IRAS\,15107+7024 (Fig. \ref{fig:15107specC} and \ref{fig:15107spec3}) cover 
the velocity range of the observed single-dish spectrum extending from 3650 to 4050 \kms with some of the weaker 
features covering the shoulders. 
The dropouts at higher velocities at the Southern location and at lower velocities at the NorthWest location appear 
unusual but they are found to be a local characteristic. They are not the result of continuum subtraction.
These differences are consistent with the presence of weak absorption structures against 
the continuum structure that do not stand out clearly in the single-dish spectrum. 

\section{Formaldehyde in Arp\,220  (IC\,4553)}

\subsection{Arp\,220 - Formaldehyde Emission}

The extragalactic formaldehyde emission from Arp\,220 was first detected using the Effelsberg telescope in 
1984 together with the absorption lines in NGC\,3628 and NGC\,3079  \citep{Baan86}. 
A representative Arecibo spectrum of the line has been presented in Figure \ref{fig:ARarp220} 
\citep[from][]{Baan93}. With a single-dish line strength of only $\sim$4.0 mJy and an isotropic luminosity 
$L_{\rm H_{2}CO} = 61\, L_\odot$, it is the strongest 
and most luminous among the known H$_2$CO MMs \citep{Araya04}. 
The formaldehyde emission is found to extend across the central molecular zones of each of the nuclei, 
while the peak of the formaldehyde spectrum at 5340 \kms is dominated by emission at the systemic 
velocity of the Western nucleus  \citep{Baan95}. 
Based on the OH MM emission at 1667 MHz and a number of high-density tracer molecular emissions 
\citep{Baan07}, the systemic velocities of the two nuclei are 5683 \kms for the Eastern nucleus and 
5365 \kms for the Western nucleus. As a result the emissions at the systemic velocity of the Eastern 
nucleus donot fall within the frequency window of these observations.

\begin{figure}
\begin{center}
\includegraphics[width=0.95\columnwidth,angle=-90]{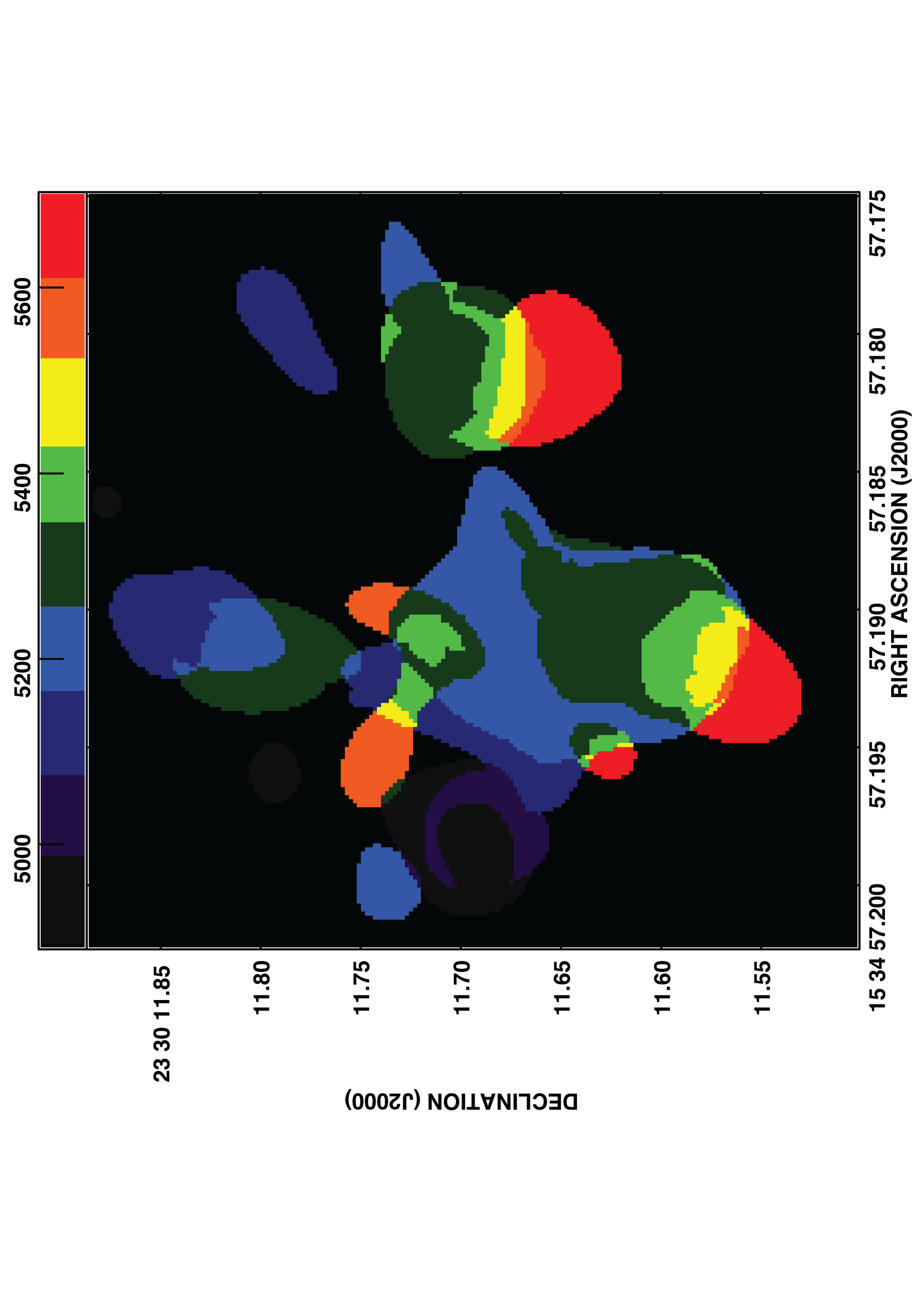}
\includegraphics[width=0.95\columnwidth]{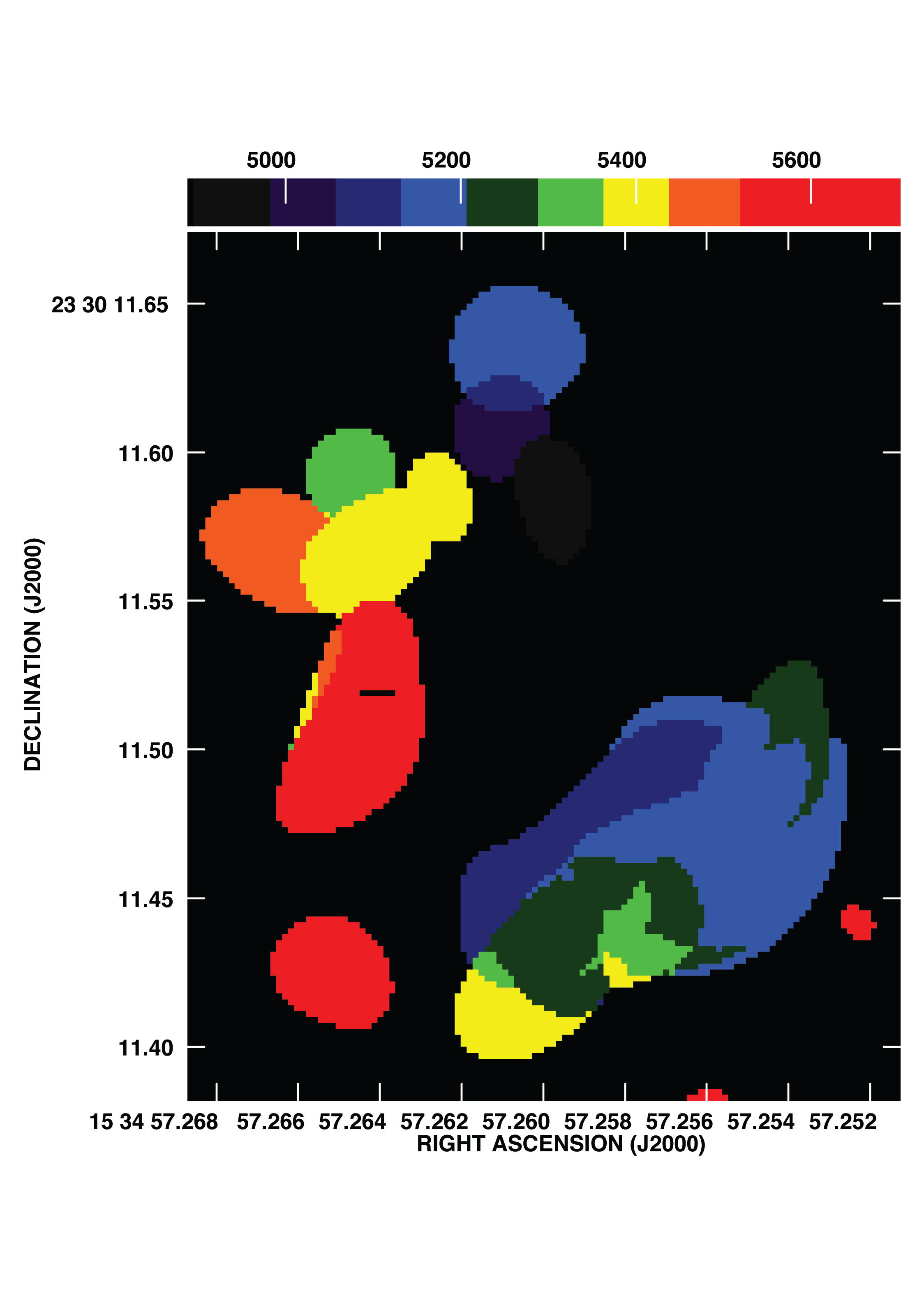}
\caption{The velocity fields at the Western nucleus (top) and Eastern nucleus (bottom) of Arp\,220. 
The emission regions in the Western nucleus possibly display a weak North - South velocity gradient and 
also the regions in the Eastern nucleus indicate a SouthWest - NorthEast gradient. 
}
\label{fig:A220EWmom1}
\end{center}
\end{figure}

\begin{figure}
\begin{center}
\includegraphics[width=0.6\columnwidth,angle=-90]{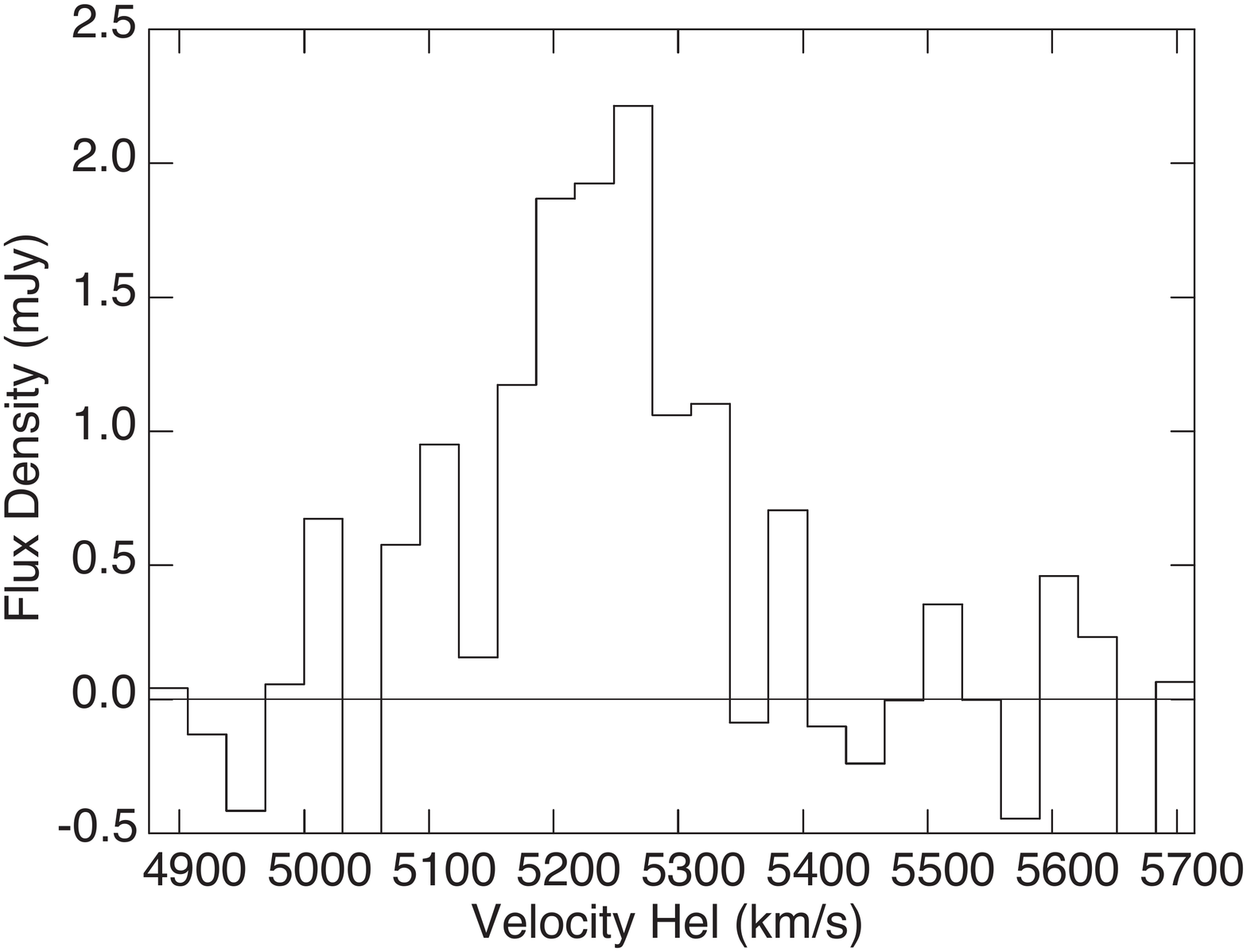}
\includegraphics[width=0.6\columnwidth,angle=-90]{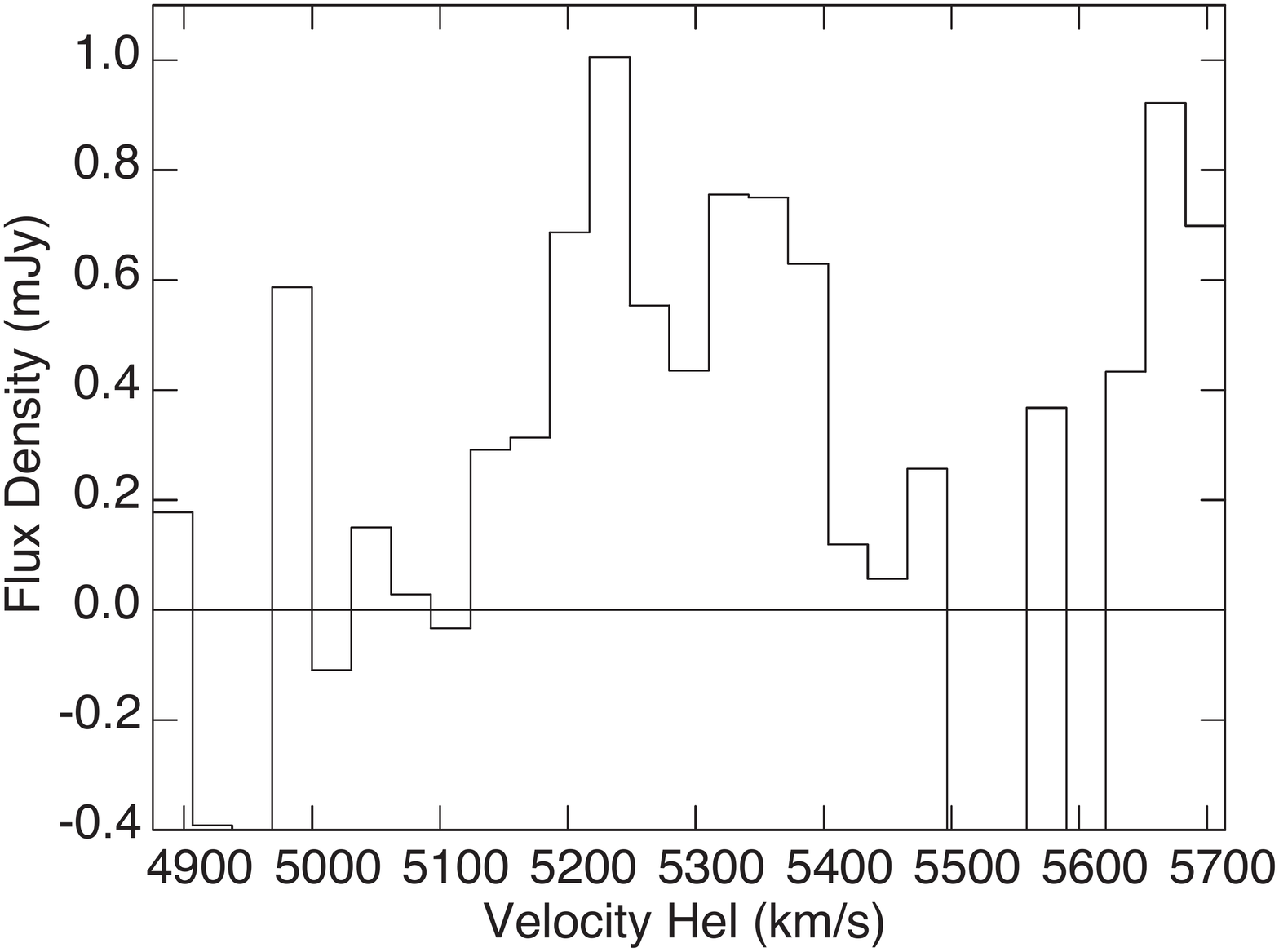}
\caption{Integrated spectra resulting from the central emission regions in the Western nucleus of Arp\,220. 
The spectra have been taken at northern part W-Centre (top) and at the southern part W-South (bottom).  
Together these emission components account for much the single-dish emission profile as presented in Fig. 
\ref{fig:ARarp220}. The structure at 5650 \kms }
\label{fig:A220Wspec1}
\end{center}
\end{figure}
\begin{figure}
\begin{center}
\includegraphics[width=0.6\columnwidth,angle=-90]{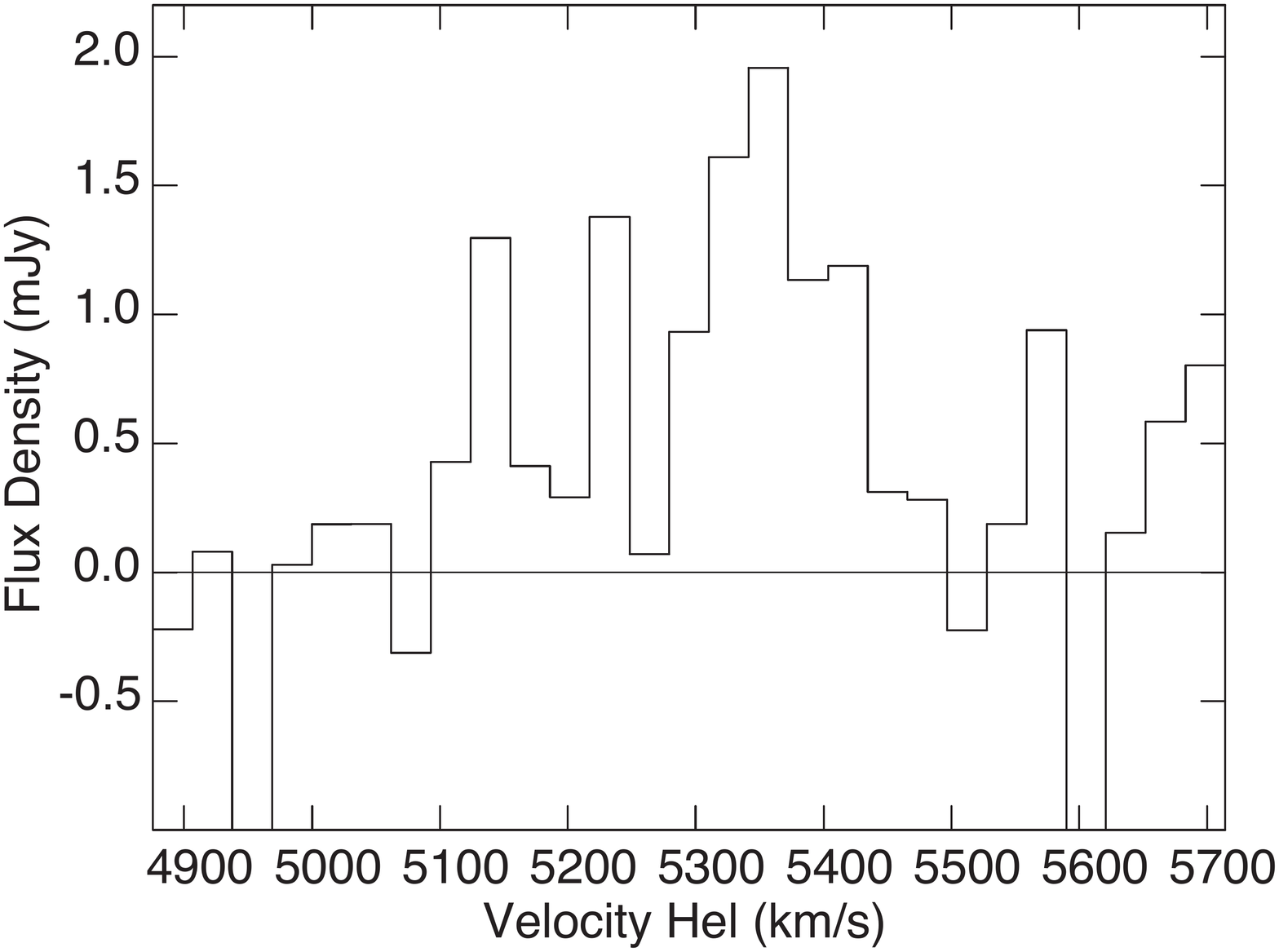}
\includegraphics[width=0.6\columnwidth,angle=-90]{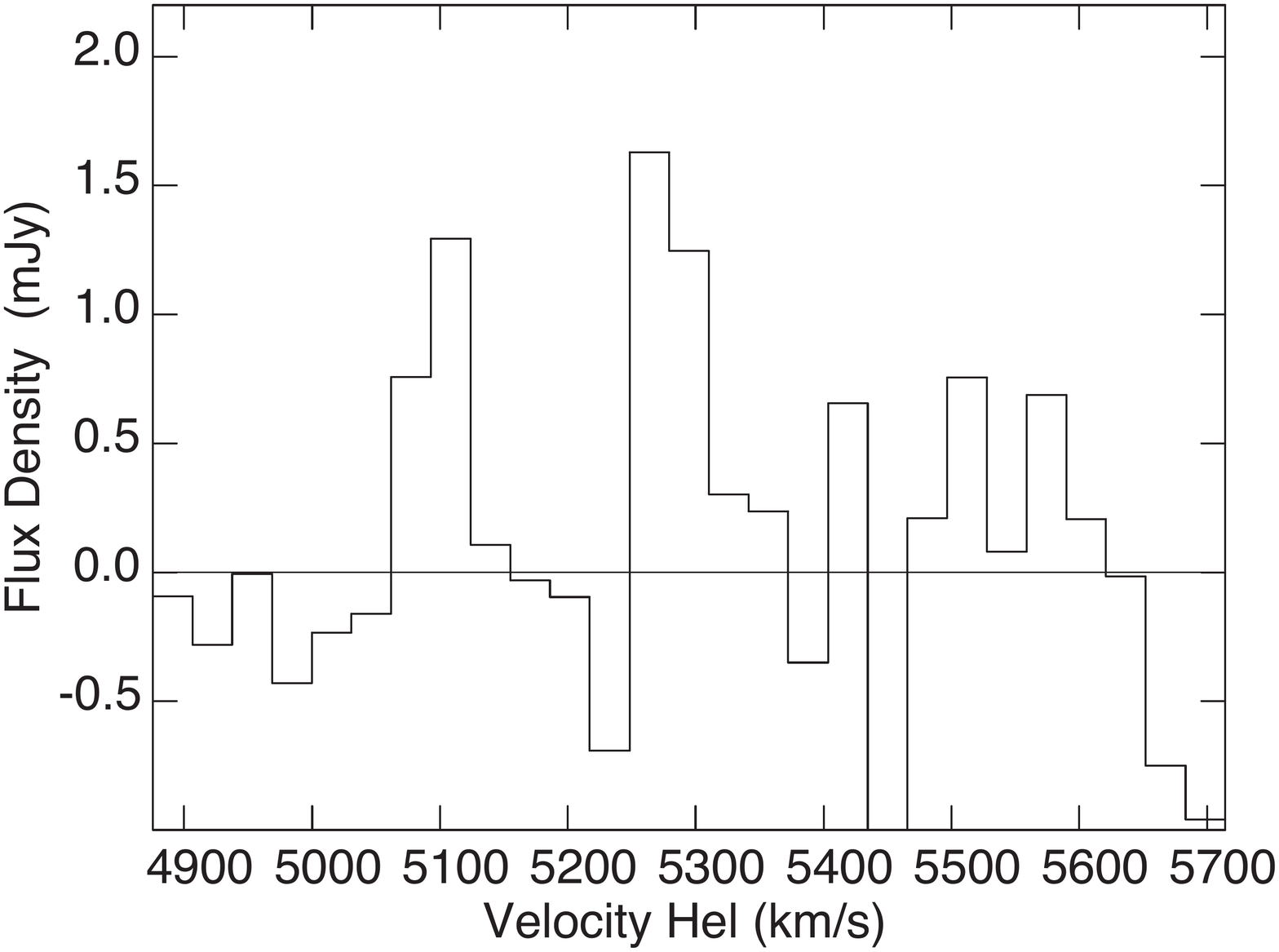}
\caption{Integrated spectra resulting from the other emission regions in the Western nucleus of Arp\,220. 
The spectra are those taken across the Western nuclear component W-West (top)
and the Northern nuclear component W-North (bottom. }
\label{fig:A220Wspec2}
\end{center}
\end{figure}
\begin{figure}
\begin{center}
\includegraphics[width=0.6\columnwidth,angle=-90]{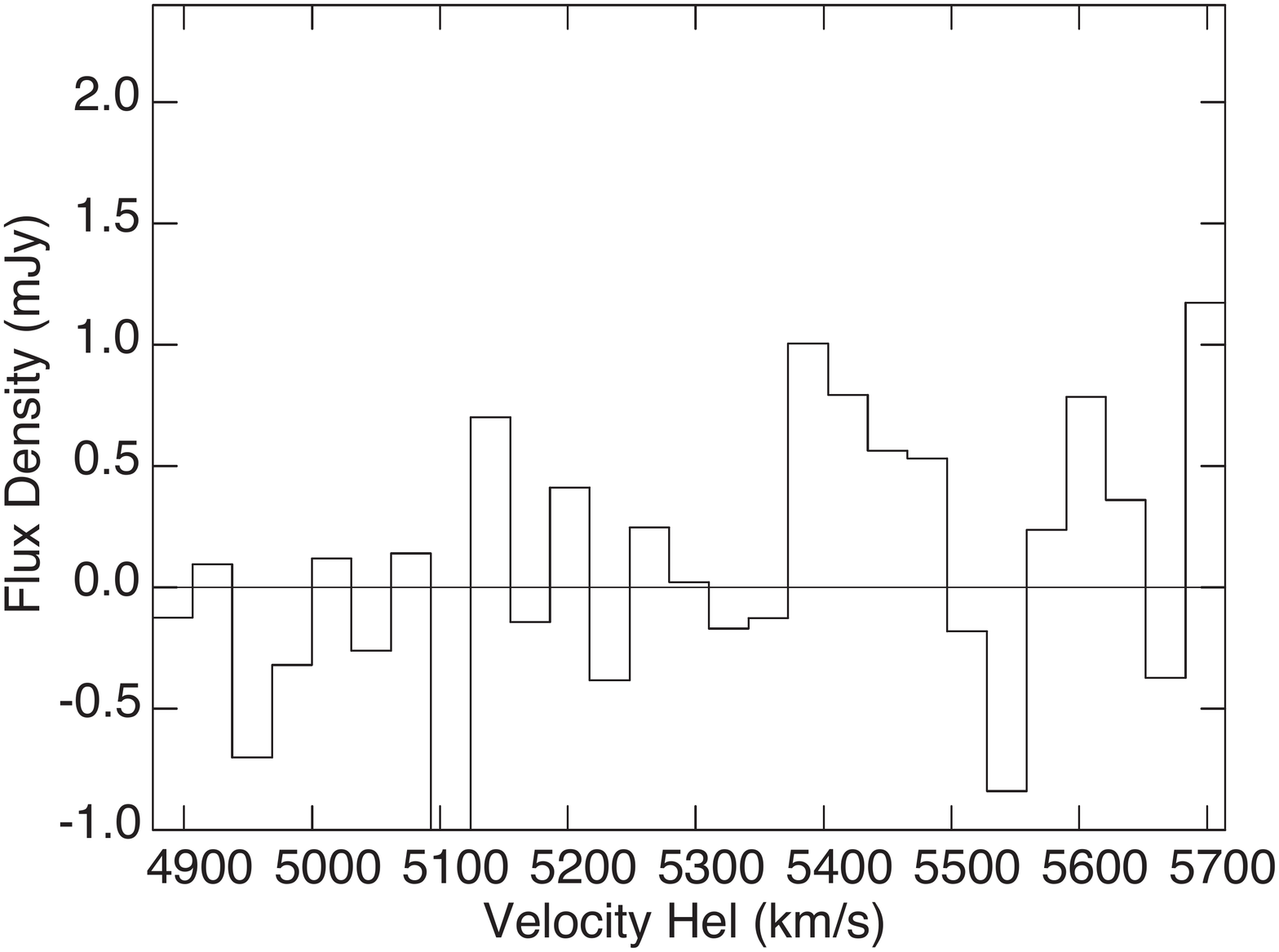}
\includegraphics[width=0.6\columnwidth,angle=-90]{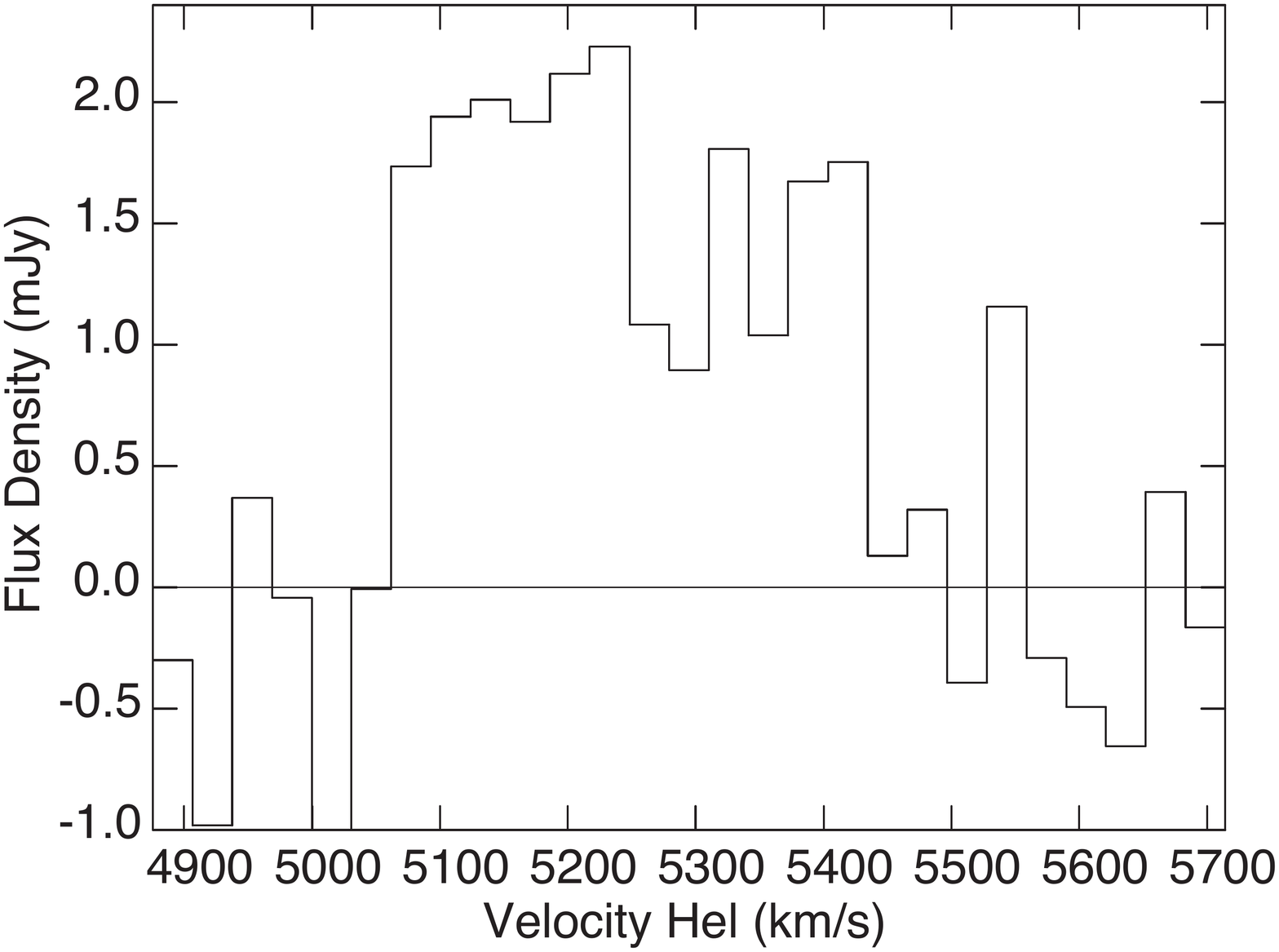}
\caption{Integrated spectra resulting from emission regions at the Eastern nucleus of Arp\,220. 
The spectra are those from the E-East region (top) and the E-West region (bottom). }
\label{fig:A220Espec}
\end{center}
\end{figure}

The formaldehyde emission structures at the nuclei of Arp\,220 are presented in Figure \ref{fig:A220comp} as a 
zeroth moment formaldehyde emission colour map superposed on the continuum contours. 
These emission regions exhibit both discrete and extended components of different 
strengths superposed on the extended continuum structure of the West and the East nuclear regions. 
The dominating central emission structure at the West nucleus is elongated by 110 pc in a North - South 
direction and is similar to the elongated structures found at the nuclei of IC\,860 and IR\,15107$+$0724. 
However, the Western nucleus shows a more complex structure than just a North-South 
nuclear disc structure. Possibly the East-West structure with a distinct (east-west) velocity gradient 
represents a superposed structural component resulting from the ongoing merger. Alternatively, these are 
star formation regions away from the disc. 
The emission region at the Eastern nucleus covers approximately 53 pc and consists of two regions in
Northeast - Southwest direction.  It should be noted that this emission occurs at the systemic velocity 
of the Western nucleus, while emission at the systemic velocity of the Eastern 
nucleus falls at the upper edge (or outside) of the observing band. 

The underlying continuum structure of the merger system Arp\,220 in Figure \ref{fig:A220comp} shows two 
isolated nuclei with peak  flux densities of the East and West nuclei of 13.38 and 30.42 mJy beam$^{-1}$). 
The peak brightness temperatures at 4.83 GHz for the eastern and western nuclei are 1.3 and 3.0 $\times 
10^{5}$ K, respectively (Table \ref{table1}), which is consistent with intense star formation. 
The (loop-like) structure that extends southward from the western nucleus to the eastern nucleus has been 
resolved out at this higher resolution \citep{Baan95,Rovilos03}. 
The extended structures at various position angles would be evidence that the star formation extends into 
pre-existing spiral arms or larger scale structures within the merging nuclei.
No evidence has been found yet for the presence of an AGN in either one of the nuclei 
\citep[][]{Smith98,Genzel98}.

\subsection{Arp\,220 - Formaldehyde Emission Structure}

The channel maps of the line emission for Arp\,220 have been presented in Figure \ref{fig:A220chan}.
The negative (single solid line) contours indicate the general noise level of each of the maps as well as 
the location of possible absorption features against the radio continuum of the source. 
The features with multiple positive contours identify the localised emission features in the maps lying mostly 
 within the contours of the radio continuum of the two nuclei. 
The features from the dominant emission regions in both nuclei above the first positive contour have been 
presented in the Moment 0 map in Figure \ref{fig:A220comp}.
The emission structure of formaldehyde in Arp\,220 is also made up of multiple (single) features that often 
fill only one spectral channel. They are also found across a large fraction of the velocity range, which does 
not fully cover the range of 5000 - 5850 \kms of the observed emission from the two nuclei (see Fig. \ref{fig:ARarp220}).

\subsection{Arp\,220 - Spectral Characteristics}

The velocity field of the emissions at the two nuclei of Arp\,220 has been presented in the first moment 
maps of Fig. \ref{fig:A220EWmom1}. Because of the 300 \kms velocity difference between the two 
nuclei, the observed emission  mostly covers the  velocity range of the West nucleus. The weaker emission 
at the systemic velocity of the Eastern nucleus enters the band only above 5600 \kms.
Both emission regions exhibit very complex velocity fields that suggest that the observed emission represents 
a superposition of compact emission regions at different radial velocities.
The emission regions at the West nucleus may show a possible North - South gradient  at 
PA = 6\degr{} and centred at V$_{west}$ = 5365 \kms across the W-North, W-Centre, and W-South regions. 
On the other hand, the W-West region shows higher velocities and the weaker W-East region shows 
lower velocities.
The emission at the East nucleus has a velocity field also centred at the V$_{west}$ velocity 
with a (possible)  Southwest - Northeast gradient at PA = 315\degr{} across the regions.  
While the orientation of these velocity gradients  are in rough agreement with those of the OH MM 
(1667 MHz) data, the velocity gradient at the West nucleus appears opposite to the local HI and OH 
gradient \citep{Rovilos03,Baan07}. 
The whole nuclear region shows a North - South gradient based on CO and HCN 
molecular data \citep{DS98,ZA08,Sakamoto08}. 
Considering that the formaldehyde material seen at the East nucleus mostly belongs to the West galaxy 
and that the observed emission has a very complex structure, no detailed information can be 
derived from the velocity structure in the moment maps.

The line emission spectra of the four prominent components in the West nucleus (Fig. \ref{fig:A220Wspec1} 
and  \ref{fig:A220Wspec2}) and of two prominent components in the East nucleus (Fig. \ref{fig:A220Espec}) show a 
complementary picture. 
The central emission regions W-Centre and W-South at the West nucleus at 5230 \kms provide the dominant peak in the 
single-dish spectrum while the emission from the western regions W-West  contributes to the 
higher-velocity shoulder of the profile. 
Region W-North in the West nucleus also contributes to the lower velocity shoulder.
The East nucleus in E-West shows multiple peaks covering a larger velocity range that contributes to both 
the low-velocity wing down to 5100 \kms and the extended high-velocity wing up to 5540 \kmss. 
This higher velocity component at the East nucleus and also the 5600 \kms structure in the W-South location 
(\ref{fig:A220Wspec1}) are consistent with emission close to the systemic 
velocity of the East nucleus as has also been seen in the OH 1667 MHz MM data \citep{Baan84, Rovilos03}.
The spectra at both nuclei as presented in Figures \ref{fig:A220Wspec1},  \ref{fig:A220Wspec2} and 
\ref{fig:A220Espec} again show the presence of localised (narrow and broad) absorption-like features.

The amplifying optical depth and the brightness temperature of the emission line features in Arp\,220 are 
presented in Table \ref{table2}.
The optical depth of the features at both nuclei varies between 0.10 and 0.29, with smaller values at locations 
with higher continuum fluxes, consistent with amplification of the continuum.
The brightness temperatures of the features vary between 2.5 and 13.4 $\times 10^4$ K. 
The high brightness temperatures of all emission components confirm masering nature of these emissions. 

\begin{table*}
\begin{center}
\caption{The line emission components}
\label{table2}
\begin{tabular}{lccccccc}
\hline
Source          &RA$^a$  & Dec$^a$ & Continuum  & Velocity  & Line     & Line  & Brightness \\
Component   &               &               &  Flux$^b$     &               &  Flux    & Optical Depth$^c$  & Temperature  T$_b$  \\
                       &   (s)     &  (\arcsec)     &   (mJy/b)      &   (km/s) &  (mJy/b)&         &    (10$^4$ K)             \\
\hline
         &           &                       &                      &               &                 &              &                 \\
\multicolumn{3}{l}{IC\,860} & & & & & \\
Centre       &  03.504 & 07.79 &  4.04 & 3490, 3640, 3830, 3990    & (3.30), 1.90, 2.80, 1.90 & (0.62), 0.34, 0.46, 0.34  & 3.01 - 5.23  \\
SouthEast  &  03.516 & 07.67 & 0.20 &  3520, 3720, 3870             & 0.71, 0.79, 1.34           & 1.26, 1.37, 1.90  & 2.23 - 4.16   \\
          &           &                       &                      &               &                 &              &                 \\
\multicolumn{3}{l}{IRAS\,15107+0724} & & & & & \\
Centre         & 13.097 &  31.92   & 6.61 &  3760, 3910                   & 1.45, 1,18                     &   0.20, 0.16            &   4.22, 3.43 \\
South          &  13.093 & 31.58   &  0.50  &  3460, 3710, 3900,4050 & 0.80, 0.95, 0.66, 0.28 & 0.95, 1.06, 0.84, 0.44 & 0.56 - 1.92 \\
NorthWest   & 13.082 &  32.03   & 0.22  &  3710, 4030                    & 0.80, 1.00                     & 1.53, 1.71  & 3.23, 4.04 \\
        &           &                       &                      &               &                 &              &                 \\
\multicolumn{3}{l}{Arp\,220 West} & & & & & \\
W-Centre   &  57.191  &  11.66  &  20.3  &  5230                              & 2.20                 &  0.10                 & 5.28 \\
W-South   &  57.191  &  11.58  &  5.12  &  5230, 5360                   & 1.03, 0.76         &   0.18, 0.14         & 3.37, 2.48 \\
W-West             &  57.181 &  11.76   &  5.19  &   5140, 5230, 5340          & 1.30, 1.38, 1.75  &  0.22, 0.24, 0.29 & 10.06, 13.43 \\
W-North            &  57.192 &  11.80   & 4.62   &   5090, 5280                    & 1.30, 1.55           &    0.28, 0.29       & 3.89, 4.64 \\
\multicolumn{3}{l}{Arp\,220 East} & & & & & \\
E-East    &  57.265 &   11.53  &  6.18         &   5430                               &  1.05                           &  0.16          &  3.14 \\
E-West    &  57.258  &  11.45   & 7.84         &   5160, 5320, 5400, 5540 & 2.20, 1.80, 1.75, 1.20 & 0.14 - 0.24 & 2.88 - 5.28 \\
\hline
\end{tabular}
\end{center}
\scriptsize{Notes: (a) Positions of IC\,860 relative to RA = 13h:15m and Dec = 24$^o$:37'. 
Positions of IRAS\,15107+0724 relative to RA = 15h:13m and Dec = 07$^o$:13'. 
Positions for Arp\,220 relative to RA = 15h:34m and Dec = 23$^o$:30'.
(b) The continuum flux density represents a mean value across the emission region.
(c) The line optical depth assumes exponential amplification of the background radio continuum.}
\end{table*}

\section{Interpreting  the Emission}

\subsection{Extragalactic Emission Structures}    
\label{sec:properties}

The formaldehyde K$_a$ = 1$_{10}$$-$1$_{11}$ data at 4.829~GHz in IC\,860, IRAS\,15107+0724, and Arp\,220 
confirms that the emission originates in regions centred on the nuclei of the galaxies and that the 
emission results from masering amplification of the radio continuum by excited foreground molecular gas. 
The dominant H$_2$CO emission arises from a central (elongated) molecular structure of size 30 - 100 pc 
in a line of sight close to the peak of the continuum source, and also from some isolated star-formation regions, 
whose projected positions are also within a region defined by the radio contours. 
The central emission regions are consistent with a clumped masering medium in the central section of an 
edge-on molecular disk causing a hierarchy of emission structures from compact high-brightness to more 
extended lower-brightness components. 
Any emission from the outer regions of the disk is undetected because of the decreasing continuum flux. 
The central regions do not show a prominent velocity gradient because of the superposition of  
dominant  point sources at different velocities superposed on a more extended structure.
An elongated emission structure is also found in Sgr B2 as a string of six single maser features covering a linear 
distance of nearly 5 pc \citep{Mehringer94,Qin08}, although there is no evidence yet of a more extended and 
diffuse component. 

The formaldehyde emission at the West nucleus of Arp\,220 coincides quite well with the presence of a 
compact circumnuclear disk as deduced from the OH MegaMaser and the CO and {\rm H I} emission.
The emission from its East nucleus may not be representative of the observed disk rotation, but it is still 
well-aligned with the larger-scale rotation structure \citep{Sakamoto99,Mundell01,Baan07}.
In the cases of  IC\,860 and IR\,15107+0724 no detailed molecular studies have yet been done. However, 
the observed formaldehyde emission structure is well aligned with the optical image of both galaxies, 
and in the case of IC\,860 it also lies perpendicular to the extended relic continuum structure.

The weaker emission components found in IC\,860 and IRAS\,15107$+$0724 represent discrete star formation regions 
in the circumnuclear environment. The emission also results from amplification of the lower background 
continuum but they require higher amplifying optical depths than those in the nuclear region (Table \ref{table2}). 
The estimated brightness temperatures of the emissions in the circumnuclear regions are slightly lower than those 
in the central regions, which gives a range for the observed temperatures  between 6 $\times$ 10$^3$ K and 
1.3 $\times$ 10$^5$ K. 
Such brightness temperatures confirm that the observed line emission results from masering activity and 
that these sources are indeed MegaMasers.

Except for the three galaxies considered here, single-dish spectra of nearby galaxies show absorption for both 
the 4.83~GHz  K$_a$=1 and the 14.5~GHz K$_a$=2 transitions \citep{Baan93,Araya04,Mangum13}. 
This would indicate that the ground state inversions are weak and possibly rare and that there is 
competition between emission and absorption in the 4.83~GHz line of most galaxies. As a result the integrated 
profiles may show evidence of absorption from extended regions, of localised emission, and possibly of localised 
re-absorption depending on the physical conditions of the intervening molecular material. 
The spectral results presented in this paper show some evidence of both narrow and broad absorption components, 
which could account for the differences when compared with the single-dish spectra (e.g. the feature at 3580 
\kms in the spectrum of IC\,860).
This spectral superposition is further supported by some recent single-dish 4.83~GHz H$_2$CO spectra of 
extragalactic sources that show clear evidence of emission components superposed on dominant absorption
 \citep[see][]{Mangum13}.

The occurrence of formaldehyde maser action remains rare in the Galaxy and in extragalactic sources.
Although there are many luminous infrared sources in the Galaxy and in extragalactic sources, only a  
small fraction of these are associated with H$_2$CO and also OH maser activity. 
Therefore, masering activity only happens during certain evolutionary stages when both the right pumping 
agent and the right molecular environment exist along a certain line of sight.
As a result, the special conditions under which H$_2$CO masering action happens hold a key for  
understanding the sequence of events during starformation and starburst processes.

\subsection{The H$_2$CO Pumping Environment}

The establishment of maser action in four extragalactic sources again raises the issue of the responsible 
pumping mechanism for formaldehyde in Galactic and extragalactic sources. 
Early suggestions for pumping the formaldehyde molecules included radiative pumping by FIR radiation 
fields \citep[e.g.,][]{Litvak69} and a pumping model using the free-free radio continuum \citep[e.g.,][]{Boland81}.  
Free-free puming can produce small level inversions but it does not explain the maser components 
discussed in this paper \citep{Baan86}, and the formaldehyde masers in Sgr\,B2 \citep{Mehringer94}.

Collisional pumping models have been suggested based on H$_2$ and electron collisions \citep{Araya07} and 
shocks \citep{Mehringer94,Hoffman07, Araya04}.  
Collisional excitation with H$_2$ and radiative excitation by the free-free radio continuum radiation from a 
nearby ultra- or hyper-compact HII region can indeed invert the H$_2$CO K$_a$ = 1$_{10}$$-$1$_{11}$ 
transition \citep{vdWalt14}. 
However, collisional pumping in shocks does not explain the simultaneous flaring of the H$_2$CO and 6.7 GHz 
Class II CH$_3$OH maser components of the Galactic source IRAS\,18566+0408 \citep{Araya10}.

The FIR radiation fields in the host galaxies of formaldehyde MM remain as the 
most viable pumping agent for extragalactic formaldehyde emitters and for Galactic sources. 
The FIR radiation fields are already found to be responsible for pumping the OH in Galactic sources and 
OH MM sources and they can also account for pumping many Class II maser transitions for Galactic 
methanol \citep{Cragg05} and naturally explains the flaring in IRAS\,18566+0408. 
A early link between formaldehyde MM activity and FIR luminosity and spectral colour temperature has been established 
based on these galaxies \citep{Baan93,Araya04}, in analogy with similar relations for OH MM sources.

\begin{figure}
\begin{center}
\includegraphics[width=1.0\columnwidth]{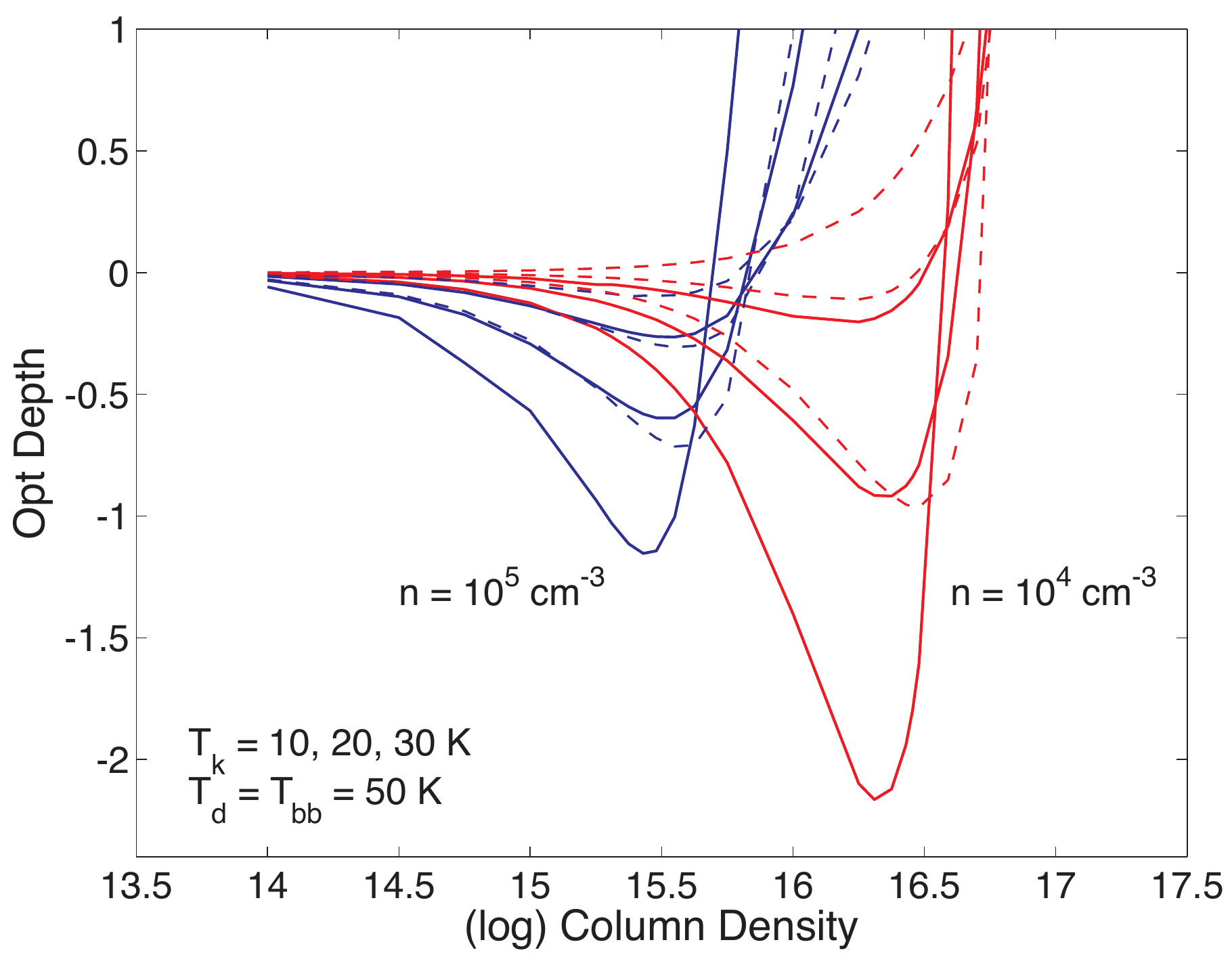}
\caption{Inversion of the formaldehyde molecules by FIR radiative pumping.
A representative dust temperature T$_{\rm d}$ = 50 K has been assumed to represent the FIR radiation field.
The optical depth curves for the 4.83~GHz K$_a$ = 1 transition are in solid lines and the curves for the 
14.5~GHz  K$_a$ = 2 transition are in dashed lines. 
The two groups of curves are for two gas densities (blue for 10$^5$ cm$^{-3}$ and red for 10$^4$ cm$^{-3}$) 
for three kinetic temperatures.  
The gains increase for lower gas densities and decrease for higher values of the kinetic temperature T$_{\rm k}$.
}
\label{fig:pumping}
\end{center}
\end{figure}

\subsection{FIR Radiative Excitation of H$_2$CO}
\label{sec:pump}

A radiative FIR pumping scenario for the formaldehyde 4.829~GHz K$_a$ = 1$_{10}$$-$1$_{11}$ transition 
requires a certain temperature range for the prevailing FIR radiation field in order to achieve population inversions. 
The estimated FIR radiation blackbody temperatures are in the range of 40 to 65 K for the three galaxies 
\citep[see][]{Gao04}, which is also the range that results in pumping of the OH molecules. 
However, inversion of the formaldehyde K$_a$ = 1 level cannot be achieved when the kinetic 
temperature (T$_k$) of the local environment is larger or equal to the blackbody temperature (T$_{\rm d}$) 
of the dust FIR radiation \citep{vdWalt14}. Therefore, extragalactic environments with coupled dust and 
kinetic temperatures will only show absorption in both transitions combined with possible thermal emission 
in the K$_a$ = 1 transition but will not show any maser activity \citep[as seen by][]{Mangum13}.

Simulations with the Radex facility \citep{vdTak07} indicate that for local kinetic temperatures lower than 
the radiative temperature, inversions in both the K$_a$ = 1 and K$_a$ = 2 transitions can be achieved by FIR 
radiative pumping.  This describes molecular environments with densities up to 10$^5$ cm$^{-3}$ where 
the kinetic and dust temperatures are yet not coupled. 
The results of the simulations depicted in Figure \ref{fig:pumping} show that the attainable (negative) optical depth 
varies strongly with density and local kinetic temperature T$_k$ and column density N(H$_2$). 
Assuming a representative dust temperature of T$_{\rm d}$ = 50 K giving the blackbody FIR radiation field, the 
optimal densities are in the range of $n = 10^4 - 10^5$ cm$^{-3}$.  
Weak or no inversions are found at both higher densities ($n = 10^6$ cm$^{-3}$), where temperature coupling 
would result, and also at lower ($n = 10^3$ cm$^{-3}$) densities. No inversion occurs when T$_k$ becomes 
equal or higher than T$_{\rm d}$. 
A maximum in the optical depth of $\tau_{4.8} = -2.2$ is found for a density of $n = 10^4$ cm$^{-3}$ and 
a temperature difference (T$_{\rm d} - $ T$_k$) = 40 K; the gains decrease with a decreasing temperature 
(T$_{\rm d} -$ T$_k$) difference. 

Similar optical depth curves are found for the optical depth in the 14.5~GHz K$_a$ = 2$_{11}$$-$2$_{12}$ transition 
except that the peak values typically scale as $\tau_{14.5} \approx 0.45\, \tau_{4.8}$ (Fig. \ref{fig:pumping}). 
However, no discernible maser emission has yet been detected in the 14.5~GHz transition in single-dish spectra 
of extragalactic sources \citep{Mangum13}. This may be explained by the much lower gains found for the 
K$_a$ = 2 transition and the disappearance of the inversion for smaller values of the (T$_{bb} -$ T$_k$) temperature 
difference.
In the parameter range where both transitions could amplify the background radio continuum, 
a radio spectral index of $\alpha$ = $-$1 and a (peak) assumed gain $\tau_{4.8}$ = $-$1.3 will generate 
an expected emission line flux in the 14.5~GHz line that is only be 9\% of the 4.8~GHz emission line strength. 
Such 14.5~GHz line emission features may indeed go unidentified in the presence of a more 
dominant absorption component.

The evaluation of the FIR pumping of the formaldehyde in a non-coupled environment show that there is 
sufficient parameter space to provide the maser action observed in the four extragalactic sources.   
The optical depths deduced for the three sources (Table \ref{table2}) lie within the range of attainable optical 
depths found with these pumping simulations. In addition, the non-coupled (dust-kinetic) temperature 
environment required for the pumping represents a plausible condition in the star-forming environments. 
The requirement of a non-coupled temperature environment may also define the transient environment that 
provides the window of opportunity for maser action and may explain the small numbers of observed 
formaldehyde masers.

\section{Conclusion}
 
The emissions in the formaldehyde ground state transition in the three galaxies, IC\,860, IRAS\,15107$+$0724, 
and Arp\,220, are extended and show a range of structural scale sizes up to 100 pc. 
The brightness temperatures of the emission features (T$_b$ = 6 x 10$^3$ - 1.3 x 10$^5$ K) 
falling within the contours of the nuclear radio continuum indicate that they are maser features 
resulting from amplification of the continuum background. 
A clumped foreground medium pumped by the FIR radiation fields will produce a hierarchy in the 
emission structure giving a combination of extended lower-brightness and compact higher-brightness 
maser components at a range of radial velocities.

The observed central emission regions  are close to the peak of the nuclear continuum and are consistent 
with them being the front section of a larger scale disk structure in the galaxies. 
The central emission regions do not exhibit clear velocity gradients resulting from disk rotation, because 
dominant high-brightness components that contribute to the emission and the complex structure 
of the molecular ISM following a nuclear merger. 
The outer parts of the disk have not yet been detected because of the outward  decrease of the continuum 
and possibly changing pumping conditions. In the case of Arp\,220 it appears that the OH emission 
from the disk in the West nucleus actually straddles the observed formaldehyde emission \citep[see][]{Rovilos03}.
In addition to the central emission regions, a number of less prominent and isolated regions have been 
found in all three sources. Most likely these are star formation regions with a higher amplifying gain 
superposed on weaker continuum components. 

The pumping of the formaldehyde emission has been associated with the FIR radiation fields in 
the (U)LIRGs, which is analogous to the pumping scenarios for OH MMs. 
The FIR radiation field can invert the formaldehyde molecular population for a density range of 
n = 10$^4$ $-$ 10$^5$ cm$^{-3}$ when the local kinetic temperature T$_k$ is lower than the 
blackbody temperature T$_{d}$ of the radiation field.  
The observed optical depths of the amplifying medium fall within the range predicted by the simulations 
of the pumping environment. 
In general, the observed optical depths are found to be relatively low, which would be consistent with the 
concept of low-gain amplification by an extended (more diffuse) and clumpy interstellar medium. 

The current data for these extragalactic sources show some evidence for localised absorption
in addition to the emission regions. The presence of absorption, localised emissions, and 
possible re-absorption in the ISM of the nuclear region may account for some of the spectral 
differences between the single-dish data and the interferometric results. The detection of localised 
emission will thus depend strongly on the beam size of the instrument and the other (unconfirmed) 
candidate formaldehyde emitters should also be studied at higher resolution. 

The more extended emission found in formaldehyde in these galaxies provide a new view of the 
medium in these sources.
In the past, maser studies have mostly concentrated on only the highest brightness components. 
 However, it is becoming increasingly clear that a large fraction of the Galactic and also extragalactic 
OH, H$_2$CO and likely H$_2$O maser emission is extended and will be at least partially resolved 
in the highest resolution experiments. 
These extended maser regions may thus provide tools to study the structure and nature of the ISM in 
these sources in combination with other diagnostic tools. 

The requirement for pumping of the formaldehyde by FIR radiation fields implies that cold 
material must still be present in the molecular ISM of the star-bursting nucleus. 
Considering that the ongoing star formation process continues to heat any regions in the ISM with 
a non-coupled dust and kinetic temperature regime, such regions will exist during a relatively small 
evolutionary time window during the relatively early stages of evolution of Galactic star formation 
and extragalactic starbursts. 
This may indicate that formaldehyde MM activity occurs at an earlier evolutionary stage than OH MM activity and this may 
also explain why only a few H$_2$CO MM have been found among the more extended OH MM sample.

For Galactic maser environments, a empirical time sequence has been suggested for maser activity 
with different flavours \citep{Ellingsen07,Breen10}, which did not (yet) include formaldehyde masers. 
A similar sequence may now also exist for the maser activity during the evolution of star-bursting nuclei 
including OH, H$_2$CO, and also Class II CH$_3$OH masers.

\section*{Acknowledgements}
This paper is written in memory of our friend James R. Cohen (1948 -- 2006) with whom this project was started. 
WAB has been supported as a Visiting Professor of the Chinese Academy of Sciences (KJZD-EW-T01). 
WAB thanks the Shanghai Astronomical Observatory staff for their hospitality during the tenure.
TA has been supported by the NSTC 973 Program (2013CB837900) and Shanghai Rising Star program. 
The e-MERLIN is a National Facility operated by the University of Manchester at Jodrell Bank Observatory 
on behalf of the UK Science and Technology Facilities Council (STFC).
This research has made use of the NASA/IPAC Extragalactic Database (NED) which is operated by the 
Jet Propulsion Laboratory, California Institute of Technology, under contract with the National Aeronautics 
and Space Administration. 
This publication makes use of data products from the Two Micron All Sky Survey, which is a joint project of the University of Massachusetts and the Infrared Processing and Analysis Center/California Institute of Technology, funded by the National Aeronautics and Space Administration and the National Science Foundation.

\end{document}